\documentclass[twocolumn]{aastex7}
\usepackage{colortbl}
\definecolor{lightgray}{gray}{0.8}

\usepackage{CJK}
\usepackage{amsmath}

\newcommand{\lr}[1]{\left(#1\right)}

\newcommand{\sigsfr}{\Sigma_\mathrm{SFR}}
\newcommand{\sigsfrTENMYR}{\Sigma_\mathrm{SFR,~10 Myr}}
\newcommand{\sigsfrHUNDREDMYR}{\Sigma_\mathrm{SFR,~100 Myr}}
\newcommand{\siggas}{\Sigma_\mathrm{gas}}
\newcommand{\sigstars}{\Sigma_\mathrm{*}}

\newcommand{\gasvdisp}{\sigma_{\mathrm{gas,z}}}
\newcommand{\tff}{t_\mathrm{ff}}

\newcommand{\tsfr}{\langle t_{\mathrm{SFR}}\rangle}

\begin{document}
\begin{CJK*}{UTF8}{gbsn}

\title{A data-driven approach for star formation parameterization using symbolic regression}
\shorttitle{A data-driven approach for star formation parameterization}

\author{Diane M. Salim (林美玲)}

\affiliation{Department of Physics and Astronomy, Rutgers University, 136 Frelinghuysen Rd, Piscataway, NJ 08854, USA}
\email[show]{diane.salim@rutgers.edu}

\author{Matthew E. Orr}
\affiliation{Department of Physics and Astronomy, Rutgers University, 136 Frelinghuysen Rd, Piscataway, NJ 08854, USA}
\affiliation{Center for Computational Astrophysics, Flatiron Institute, 162 Fifth Avenue, New York, NY 10010, USA}
\email{matt.orr@rutgers.edu}

\author{Blakesley Burkhart}
\affiliation{Department of Physics and Astronomy, Rutgers University, 136 Frelinghuysen Rd, Piscataway, NJ 08854, USA}
\affiliation{Center for Computational Astrophysics, Flatiron Institute, 162 Fifth Avenue, New York, NY 10010, USA}
\email{b.burkhart@physics.rutgers.edu}

\author{Rachel S. Somerville}
\affiliation{Center for Computational Astrophysics, Flatiron Institute, 162 Fifth Avenue, New York, NY 10010, USA}
\email{rsomerville@flatironinstitute.org}

\author{Miles Cramner}
\affiliation{Institute of Astronomy \& Department of Applied Mathematics and Theoretical Physics \& Kavli Institute for Cosmology, University of Cambridge, Madingley Road, Cambridge CB3 0HA, UK}
\email{mc2473@cam.ac.uk}
\begin{abstract}

Star formation (SF) in the interstellar medium (ISM) is fundamental to understanding galaxy evolution and planet formation. However, efforts to develop closed-form analytic expressions that link SF with key influencing physical variables, such as gas density and turbulence, remain challenging.
In this work, we leverage recent advancements in machine learning (ML) and use symbolic regression (SR) techniques to produce the first data-driven, ML-discovered analytic expressions for SF using the publicly available FIRE-2 simulation suites. Employing a pipeline based on training the genetic algorithm of SR from an open software package called \textsc{PySR}, in tandem with a custom loss function and a model selection technique which compares candidate equations to analytic approaches to describing SF, we produce symbolic representations of a predictive model for the star formation rate surface density ($\sigsfr$) averaged over both 10 Myr and 100 Myr based on eight extracted variables from FIRE-2 galaxies. The resulting model that PySR finds best describes SF, on both averaging timescales, features equations that incorporates the surface density of gas, $\siggas$, the velocity dispersion of gas $\sigma_{\mathrm{gas,~z}}$ and the surface density of stars $\sigstars$. Furthermore, we find that the equations found for the longer SFR timescale all converge to a scaling-relation-like equation, all of which also closely capture the intrinsic physical scatter of the data within the Kennicutt-Schmidt (KS) plane. This observed convergence to physically interpretable scaling relations at longer SFR timescales demonstrates that our method successfully identifies robust physical relationships rather than fitting to stochastic fluctuations.

\end{abstract}



\section{Introduction}\label{sec:intro}


Star formation (SF) remains one of the major unsolved problems in galactic astrophysics. While there are many challenges to modeling star formation, including understanding the initial mass function of stars, a major uncertainty is how galactic-scale processes affect the rate of star formation per unit mass of interstellar gas \citep{KrumholzReview2014,MyersEtAl2014}.
A general theory for star formation is therefore crucial for thoroughly understanding  the physics fundamentally governing galaxies and their substructures at different scales. A consensus on such a general SF theory has been elusive, due to the plethora of competing complex physical processes and the vast range of scales involved \citep{Klessen2011, OffnerEtAl2013}. From observations, there exists a clear interdependence between star formation and the dynamics of the surrounding cold dense gas. This is described empirically by the  Kennicutt-Schmidt SF power-law relation \citep{Kennicutt:1998aa}, which relates the column density of the star formation rate, $\Sigma_{\mathrm{SFR}}$, to the gas column density, $\siggas$ (henceforth K98):

\begin{align}
    \Sigma_{\mathrm{SFR}} \propto \Sigma_{\mathrm{gas}} ^{1.4}~. \label{eq:K98}
\end{align}
This empirical relation has been extensively compared to observations \citep{Schmidt1959, Bigiel:2008aa, LeroyEtAl2008, DaddiEtAl2010HighZDisks, SchrubaEtAl2011, RenaudEtAl2012, Kennicutt:2012aa}.
This scaling relation holds such historical significance that this plane in the logarithmic scale is referred to as the Kennicutt-Schmidt (KS) plane. However, significant scatter has also been observed to exist in the KS plane, such that $\sigsfr$ can vary by more than an order of magnitude for any given input $\siggas$ \citep[e.g.,][]{WuEtAl2010, LadaLombardiAlves2010, GutermuthEtAl2011, HeidermanEtAl2010, Federrath2013sflaw}.

Subsequent analytic models attempting to better capture the physics of $\sigsfr$ and reduce the scatter between this variable and its descriptor have typically approached star formation from a "bottom-up" or a "top-down"  viewpoint.
The "bottom-up" perspective suggests that SF is a local process that does not depend on galactic properties \citep{KrumholzMcKee2005, KrumholzMcKeeTumlinson2009, LadaEtAl2012, RenaudEtAl2012, Federrath2013sflaw, Krumholz2013SFMoleculePoorGalaxies, SalimEtAl2015, Burkhart:2018aa}.
Evidence for this includes the observed lack of correlation between star formation rates and galactic parameters like Toomre-Q \citep{LeroyEtAl2008, LeroyEtAl2013} and the little variation in molecular gas depletion times on kiloparsec scales in nearby galaxies \citep{LeroyEtAl2013}. Furthermore, the consistency of ratio between the rate in which gas is converted to stars to the free fall time (i.e., constant efficiency per free fall time) across the entire scale of the star-forming systems, from individual clouds to starburst galaxies \citep{KrumholzDekelMcKee2012}, also supports this theory, although this rate may vary slightly depending on cloud local variables, such as the Mach number of turbulence \citep{Federrath2013sflaw}. In such bottom-up models, the global star formation rate in galaxies is primarily determined by the sum of the star formation activity in individual clouds, with little influence from the larger galactic environment.
By encompassing the physics of star formation feedback on small scales, this suggests that $\Sigma_{\mathrm{SFR}}$ is regulated by that of local variables \citep{SunEtAl2020, SunEtAl2022}.

However, these "bottom-up" models fail to express how local molecular cloud properties, such as the turbulence velocity dispersion, are set,  do not address the stability of the galaxy, nor provide an explanation for long-term gas fueling which is important for long-lived molecular clouds \citet{KrumholzBurkhart2018GalacticDiscsModel, JeffresonEtAl2021}. They thus do not directly confront issues of the \textit{dynamical} state of the gas in galaxies. To account for this, the `top-down' approaches suggests that the star formation rates in galaxies are influenced by galactic environment \citep{ThompsonQuataertMurray2005, OstrikerMcKeeLeroy2010, HopkinsQuataertMurray2011, FaucherGiguere13, KrumholzBurkhart2018GalacticDiscsModel}, supported by recent observation and numerical works which highlight the potential significance of galactic dynamics on molecular cloud evolution and large-scale shearing motions \citep[e.g.,][]{HughesEtAl2013, ColomboEtAl2014, MeidtEtAl2013, SuwannajakTanLeroy2014, KrumholzBurkhart2018GalacticDiscsModel, JeffresonEtAl2021}.
In the  "top-down" model approach, the SFR is typically set by a balance between energy injection and dissipation within a galactic disk, along with vertical hydrostatic balance, thus explaining the origins of SF scaling relations as a result of gas dynamics and migration in galaxies \citep[e.g.,][]{ OstrikerShetty2011, FaucherGiguere13, OstrikerKim2022}.
Yet while top-down approaches are advantageous as they are established upon simple physical conditions, they do not necessarily predict local variance due to stochastic processes and struggle when feedback and dynamical timescales are inverted \citep{SparreEtAl2015}.

There is no agreed upon physical model for the KS relation, nor is it clear which philosophical approach (i.e., `top-down' or `bottom-up') is most appropriate to describe SF. The model chosen for SF is important in the context of application as subgrid models of large volume or coarse resolution simulations, which represent the physical processes occurring at scales smaller than the grid resolution that can be resolved directly by the simulation. SF  ultimately drives galaxy growth and evolution, thus the final outcome of a simulation that leverages an SF subgrid model hinges fundamentally on an informed choice for its SF prescription. However, many such large-volume simulations such as Illustris \citep{VogelsbergerEtAl2014Illustris}, its successor IllustrisTNG \citep{PillepichEtAl2018IllustrisTNG} and the Evolution and Assembly of GaLaxies and their Environments' (EAGLE) \citep{CrainEtAl2015EAGLE} simulation simply adopt models based on the aforementioned \citet{Kennicutt:1998aa} empirical scaling of gas density with SFR density, above a density threshold, paired with a standard \citet{Chabrier2003} initial mass function (IMF). Subgrid models for SF also play a central role in the machinery behind semi-analytic models (SAMs).
These are computationally inexpensive models for galaxy evolution, embedded within the framework of structure formation in the cold dark matter (CDM) model, which can predict a wide range of observable galaxy properties.
Evolution is dictated by approximate or analytic approaches to associated physical processes such as gas cooling, star formation and supernova feedback \citep{Baugh2006Review, Benson2010Review}. Once again, a simple KS-law-like prescription is often adopted in SAMs  \citep[e.g.][]{ColeEtAl2000SAM, SomervilleDave2015, NaabOstriker2017}.

In view of the need to reconsider the physics of SF that go into the subgrid recipes for both large-volume simulations and SAMs, coupled with the disagreement of the mathematical forms between the two main categories of analytic SF models, in this work, we take inspiration from the recent proliferation of data-driven and machine-learning (ML) based algorithms to find an automated, agnostic mapping between features in high-dimensional datasets. The target variable in this work is the column density of star formation, $\sigsfr$, and the input variables are those on which the target variable depend, such as $\siggas$. Here, the terms "feature" and "variable" will be used interchangeably. Neural networks and gradient boosting frameworks such as XGBoost \citep{Chen2016XGBoost} have been very successful supervised learning methods for regressing in complex feature spaces, but as deeper networks are required for more complicated problems involving bigger datasets and higher dimensional spaces, these mappings between the input and output features become increasingly difficult to decipher.

As models described by mathematical equations are inherently interpretable and comparable to already existing analytic models, in this work, we explore using an ML technique known as "symbolic regression" (SR) to search for equations in a multi-dimensional dataset. In particular, we employ an open-source SR library named \textsc{PySR} \citep{Cranmer2023PySR}, which has already been applied to a range of scientific fields, with some applications in astronomy such as finding cosmological parameters from halo and galaxy catalogs \citep{ShaoEtAl2023} and black hole mass scaling relations \citep{DavisJin2023}, but has not yet been leveraged to search for a universal star formation relation, which we will pursue in this work. We do so by applying \textsc{PySR}, with a custom loss function, to the FIRE-2 (Feedback in Realistic Environments) cosmological zoom-in simulations, and developing a novel pipeline for choosing the equations that best describe the data based on comparison to previously developed analytic functions for star formation. ML applications using the FIRE collaboration simulations have included predicting high-resolution baryon fields \citep{BernardiniEtAl2021EMBER} from the FIREbox dark matter simulations \citep{FeldmannEtAl2023FIREbox} using a generative adversarial network (GAN), as well as classification of star clusters in the FIRE-2 Latte suite \citep{OttoEtAl2022}, but thus far no FIRE suite of simulations have yet to be applied to symbolic regression SF problems.

In this work, we thus leverage SR to produce the first data-driven parameterizations of star formation on $\sim$kpc scales. In Section~\ref{sec:methods}, we describe this method and the data we use for the symbolic regression training, the results of which are presented in Section~\ref{sec:results_and_analysis}. We discuss these results in Section~\ref{sec:discussion} and in Section~\ref{sec:conclusions}, we conclude our findings.

\section{Methodology} \label{sec:methods}

\subsection{Finding equations using Machine Learning}

\begin{figure*}
    \includegraphics[width=\linewidth]{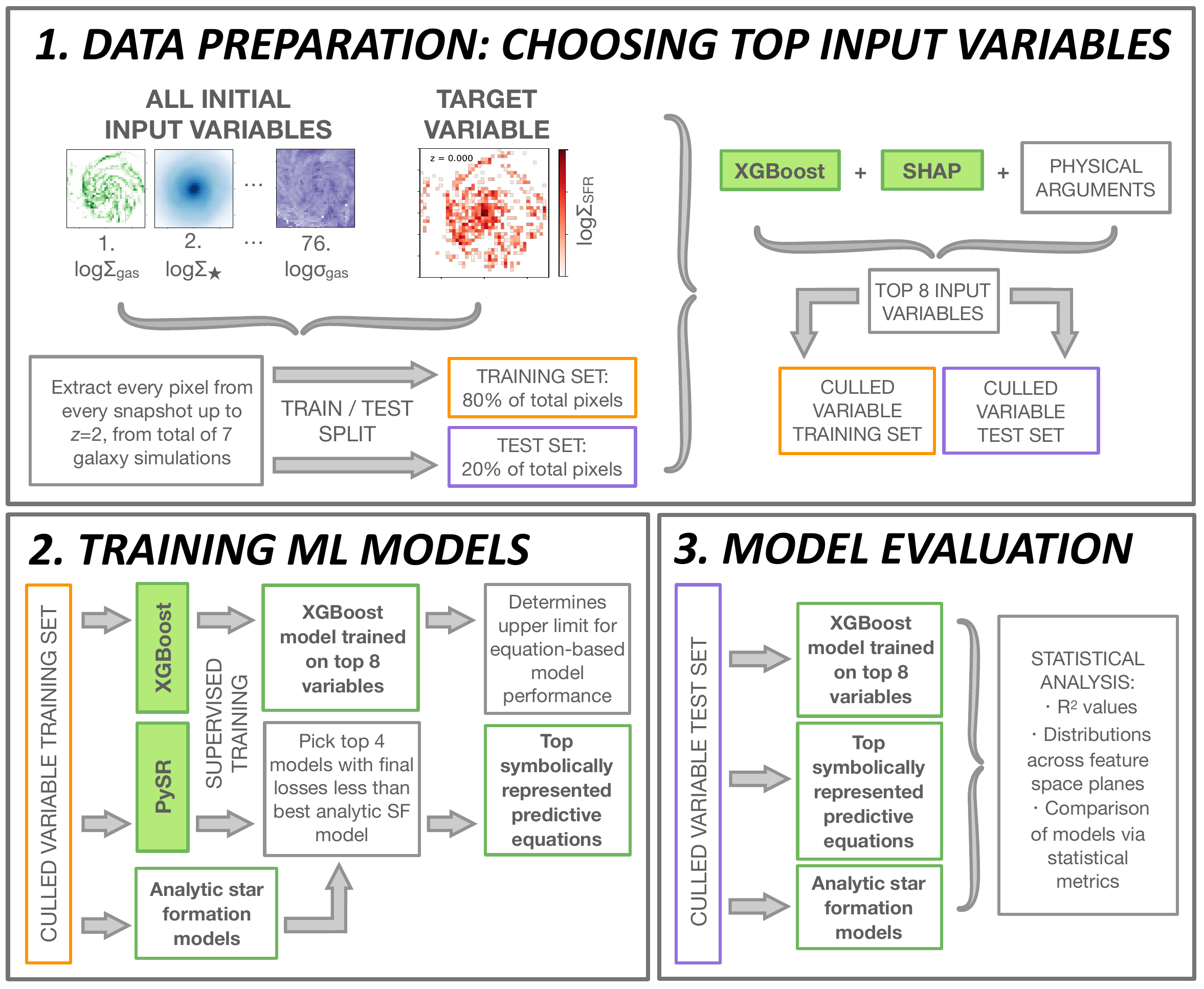}
    \caption{Chart showing how data is manipulated and flows through our procedure. In all panels, filled green boxes represent a machine learning-based method or algorithm that is utilized, and uncolored green outlined boxes represent a predictive model for our target variable, star formation rate surface density. The top panel shows how our initial raw data is prepared and filtered to determine the variables with which to run our experiments. The bottom left panel depicts supervised training being applied to the filtered variable training set in order to attain predictive models for the target variable. The bottom right panel shows how all the predictive models in this study are compared by being applied to the filtered variable test set.}
\label{fig:methodology_flowchart}
\end{figure*}
Figure~\ref{fig:methodology_flowchart} shows an overview of the experiment-pipeline in this paper, from the input data to the final analysis of found star formation scaling equations. The pipeline can be broken down into three main sub-methods. The first procedure is the preparation of our input data for training, by cutting down the 76 candidate input variables from the simulations to just 8. The culled variable training set is then used to train ML models in the subsequent procedure. The culled variable test set is then applied to the trained models in the final procedure. Below we will describe each sub-method in detail.

\subsubsection{Data Preparation: Choosing Top Input Variables}\label{subsubsec:data_prep}
The first sub-method in our pipeline is the procedure by which we choose our top input variables. We do this to streamline training using symbolic regression in subsequent steps. The initial dataset had 76 candidate input variables, the maps of which we extract on a pixel scale and perform a train/test split of 80$\%$ training and 20$\%$ test data. In order to decide on the input parameters which will have the most influence on our target variables, we first apply XGBoost \citep{Chen2016XGBoost}, which stands for "Extreme Gradient Boosting" and uses gradient boosted decisions trees for regression, classification and ranking problems. For an in-depth explanation regarding the workings of this algorithm please refer to \citet{Chen2016XGBoost}. In this study we utilize it for regression in order to find a mapping between pixel-scale characteristics of the galaxies as input parameters and the surface density of star formation as the output parameter. We train XGBoost with all 76 candidate input variables on the training set. We conduct our training in the logarithm of 10 space due to the large dynamic range of our target variable and all our feature variables. The resultant trained model, along with the test data, is fed through a SHapley Additive exPlanations (SHAP) \citep{Lundberg2017SHAP} analysis to explore which of the input variables are most influential in describing our target variable. This analysis assigns a SHAP value to every pixel, the calculation of which are based on game theory. Positive SHAP values indicate that the feature is positively impacting, or increasing, the model's prediction of the target variable, whilst negative values indicate negative impact, or a decrease in the prediction of that target variable. For example in our experiment, if dense regions with high $\siggas$ values yield positive SHAP values this indicates that the XGBoost model predicts high $\sigsfr$ values for those corresponding regions. Likewise, less dense regions yielding negative SHAP values show that the model predicts that decreasing the density of the gas corresponds to a decrease in $\sigsfr$. The magnitude of the SHAP value is a sign of the strength of the effect. From this analysis, we are able to attain a ranking of which input parameters are most influential to predicting our target variable. However, SHAP values do not explicitly measure the accuracy of a model. We pair this ranking with some physical arguments, such as the fact that the column density of star formation is traced by H$\alpha$, thus the latter would not be a suitable parameter to investigate as its correlation to the target variable would be trivial. In doing so we are able to choose in a data-driven manner eight top input variables: the surface density of neutral gas $\siggas$ and that of stars $\sigstars$, the vertical ($z$-axis) velocity dispersion of gas $\sigma_{\mathrm{gas,z}}$ and that of stars $\sigma_{\mathrm{*,z}}$, the inverse dynamical time $\Omega_{\mathrm{dyn}}$, the local gas fraction\footnote{The gas fraction is degenerate with including $\siggas$ and $\sigstars$, however, we chose to include it separately as \textsc{PySR} might have found a more compact mathematical representation for SFR.} $f_\mathrm{gas}=\siggas/(\siggas + \sigstars)$, and finally the velocity of the gas in the $\phi$ and $z$ directions, $V_\mathrm{gas,\phi}$ and $V_\mathrm{gas,z}$, respectively. For specific details regarding the origins of these input variables and the FIRE simulations please refer to Section~\ref{subsec:FIREsims} below.

\subsubsection{Training Machine Learning Models}
We apply the symbolic regression (SR) equation search algorithm \textsc{PySR} \citep{Cranmer2023PySR} in order to explore the range of best equations that are representative of the dataset. \textsc{PySR} is a genetic algorithm whereby equations are represented as a graph, with the nodes in the graph corresponding to a term or operator in the equation. With each iteration of genetic algorithms, a phenomenon occurs to the graph akin to evolution after a generation. Examples include a random mutation, whereby a node changes its operator, or a crossover, in which two separate branches within the graph can switch. In this way, new "generations" of equations are constructed. The number of nodes in a given equation defines its "complexity". There is not only one solution but a population, which refers to a collection of candidate solutions, and this population is evolved over multiple generations to find the most optimal solution. \textsc{PySR} in particular is multi-population evolutionary algorithm, meaning that multiple populations are evolved asynchronously, parallelized over workers. After the populations have evolved after a generation, a "migration" occurs in which a small fraction of equations are replaced with "migrating" equations from either other populations or a set of top equations that have been found across all populations during training thus far. This sequence of evolving a population then subsequently "migrating" a few of the resulting equations is referred to as an iteration. For more details regarding the algorithm and optimization methods of \textsc{PySR} please refer to \citet{Cranmer2023PySR}. In our particular experiment, we call the main equation-finding algorithm function \textsc{PySRRegressor} set with the number of populations running to be 128 (as we are running on 64 cores and the optimal value is $2\cdot n_\mathrm{cores}$) and the number of total mutations to run, per 10 samples of the population, per iteration to be 5000. This is a relatively large value for this parameter and results in a slower algorithm search but more efficient use of computer cores. We choose to utilize batching, which compares population members on small batches during evolution, and set the batch size to 256. Furthermore, because the \textsc{PySR} software does not inherently save loss values after every iteration, and we desired confirmation of convergence of solutions by the end of training which can be checked via the evolution of loss values as training progresses (i.e., a loss curve), we set the regressor to perform a "warm start", whereby the model optimizations continue from where the last call to fit had finished. We train for just two iterations at a time, calculating and saving the loss values after each of these cycles, which, in this study, we refer to as an "epoch". Concerning our search space, we constrain terms containing our input features to be linear combinations of each other (i.e., found equations can not contain direct products of input features) due to their logarithmic nature, as well as in order to improve interpretability. Exponential and logarithm of 10 operators are also permitted, but no such functions can be expressed inside any said function.

In short, on the face value, the final output of \textsc{PySR} are functional equations $\Psi$ predicting a relation between the target variable and the input variables. Therefore the final predictions of the target variable can be characterized as:
\begin{align}
   \widehat{\mathbf{y}} = \Psi(\mathbf{x}).
\end{align}

We train \textsc{PySR} for 4000 epochs with the aforementioned input variables , thus for this experiment, we have:
\begin{equation}
    \mathbf{x}:=(\Sigma_{\mathrm{gas}},~\Omega_{\mathrm{dyn}},~\sigma_{\mathrm{gas}},~f_g,~\Sigma_{*},~\sigma_{\mathrm{*}},~V_\mathrm{gas,\phi},~V_\mathrm{gas,z} ) .
\end{equation}

Note that once again, we train in the logarithm of 10 space due to the large dynamic range of all the variables. The prediction given by $\widehat{\mathbf{y}}$ of the found equation is compared to the values at the same location in the native FIRE simulation ${\mathbf{y}}$ via loss constructed as the summation of standard mean squared error (MSE) loss and the quantile loss. The MSE loss, also known as an $l_2$ regularized loss or prediction loss, is given by:
\begin{align}
  \mathcal{L}_{\mathcal{P}} = \dfrac{1}{N}\sum_{j=1}^{N}{\lr{\mathbf{y}_{j} - \widehat{\mathbf{y}}_{j}}\cdot\lr{\mathbf{y}_{j} - \widehat{\mathbf{y}}_{j}}} , \label{eq:pred_loss}
\end{align}
where $N$ is the number of defined points, i.e., pixels, in the simulations.
We also utilize a quantile loss for quantiles of 0.1, 0.25, 0.5, 0.75 and 0.9. The quantile loss is defined as:
\begin{align}
    \mathcal{L}_\mathcal{Q} = \dfrac{1}{N_{\alpha}}\sum_{i=1}^{N_{\alpha}}{\left[ \begin{cases}
                        \alpha_{i} (\mathbf{y} - \widehat{\mathbf{y}}) & \text{if } \mathbf{y} \geq \widehat{\mathbf{y}}, \\
                        (1 - \alpha_{i}) (\widehat{\mathbf{y}} - \mathbf{y} ) & \text{if } \mathbf{y} < \widehat{\mathbf{y}},
                      \end{cases}\right]}
\end{align}\label{eq:quantile_loss}
for the $N_{\alpha}=5$ quantiles we are considering. We choose to incorporate the quantile loss as it predicts for an interval rather than a single value, thus has an ability to provide information about the entire distribution of a target variable, not just the mean. This allows for more robust predictions, especially when the dataset being considered contains many outliers or skewed distributions \citep{TervenEtAl2023, WangEtAl2022}. We choose the total loss to be the sum of these constituent losses with equal weighting:
\begin{align}
     \mathcal{L} = \mathcal{L}_\mathcal{P} + \mathcal{L}_\mathcal{Q}. \label{eq:LOSS_EQN}
\end{align}

Symbolic regression algorithms like \textsc{PySR} result in not a single solution, but an entire Pareto-front, which is the output of a range of equations with different complexities trading-off with accuracies \citep{Cranmer2023PySR, KommendaEtAl2021}. Therefore, one of the challenges of utilizing these algorithms is finding an optimization to model selection in order to single out the equation that, indeed, best represents the data. The standard statistical approach of taking the model with the highest $R^2$ metric is perilous and biased in this scenario, because naturally the more complexity an equation is allowed, the more degrees of freedom it is afforded to fit (and potentially overfit) the data. A common and straightforward method is to calculate the "score", a value that assesses the largest ratio between the change in loss and change in complexity on the Pareto front, and choosing the equation with the highest of this metric is the default method used by \textsc{PySR} to indicate its best model selection. In this study, we, too, utilize the "scores" of the resulting equations to rank them, but our "top selected equations", of which we choose four per experiment, additionally need to exhibit final training losses less than that of the best analytic model applied to the data set. The range of analytic models investigated is described below in Subsection~\ref{subsec:comparable_models}. We choose to adopt this method because the physics-motivated analytic models for SF give a physical justification for the complexity expected out of a parameterization for the target variable, and choosing via a sole ranking of the top "score" is likely to favor less complex equations which may capture the first order influences for SF but may be reductive of the many nuanced feedback processes that ultimately drive this phenomenon.

\subsection{Training dataset: FIRE-2 simulations}\label{subsec:FIREsims}

\begin{figure*}
    \includegraphics[width=\linewidth]{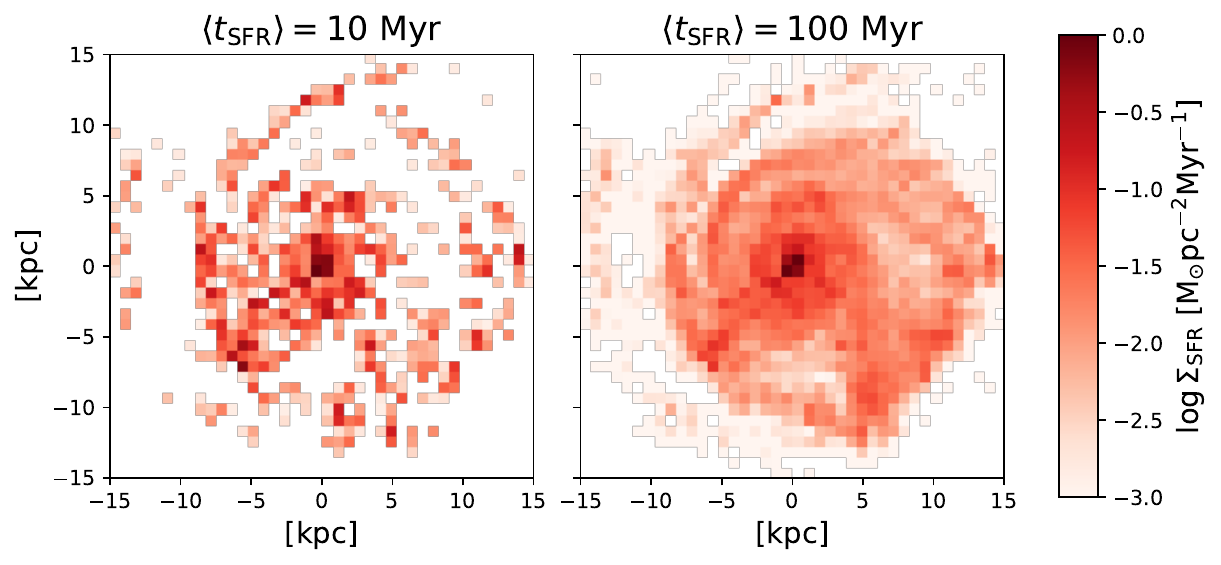}
    \caption{Star formation rate surface density ($\sigsfr$) maps of galaxy "m12b" at redshift $z=0$ with 750 pc pixel size, averaged over 10 Myr (\textit{left panel}) and 100 Myr (\textit{right panel}).}
\label{fig:m12b_ZSFR}
\end{figure*}

In this study we make use of $z \approx 2-0$ snapshots from the public data release of a subset of the FIRE-2 (Feedback in Realistic Environments) cosmological zoom-in simulations, called the Latte suite, which includes seven galaxies \citep{WetzelEtAl2023FIREDataRelease}. These isolated MW/M31-mass galaxies, first introduced in \citet{WetzelEtAl2016} and \citet{HopkinsEtAl2018FIRE2}, were run with the GIZMO code \citep{Hopkins2015GIZMO} using the mesh-free finite mass Lagrangian Godunov (MFM) mode adopting a standard \citet{PlanckCollaboration2014}-$\Lambda$CDM cosmology.

These simulations are well-suited for testing a method to extract new laws (equations) for star formation on sub-kpc scales because the SF prescription in the FIRE-2 simulations is physically motivated with minimal free parameters set by hand, the details of which are reported in \citet{HopkinsEtAl2018FIRE2}. To summarize here, for SF to occur, gas must meet the criteria that it is dense ($n>10^3 \mathrm{~cm}^{-3}$), molecular ($f_{\rm H_2} > 0.5$ as per the \citet{KrumholzGnedin2011} empirical fit for molecular gas fractions as a function of local gas column density), self-gravitating (viral parameter $\alpha_{\mathrm{vir}} < 1$), and finally, Jeans-unstable below the resolution scale. Meeting all four criteria allows gas elements to be converted into star particles at a rate $\dot\rho_\star = f_{\rm H_2} \rho_{\rm gas}/t_{\rm ff}$, after which it is treated as a single stellar population with known age, metallicity, and mass. Feedback from supernovae (Type Ia and II), stellar mass loss (OB/AGB-star winds), photoionization and photoelectric heating, and radiation pressure is thereafter explicitly followed, with all feedback quantities taken from the standard \citet{LeithererEtAl1999STARBURST99} STARBURST99 stellar population models with an assumed \citet{Kroupa2001} IMF. Therefore, unlike large-volume simulations that encompass all small-scale physics to a subgrid model that is often chosen to be the observed KS relation, in the FIRE-2 simulations no specific empirical star formation law or scaling has been put in as a prescription \textit{by hand} \citep[see, e.g.,][for an early exploration of the emergence of Kennicutt-Schmidt using FIRE-1]{Orr2018FIREKSlaw}. They are, however, rich with physics stemming from the aforementioned myriad sources of feedback, which all together produce a highly multi-phase and dynamic ISM.

For our analyses, we generate spatially resolved maps of the simulations in an `IFU-like' fashion by aligning the galaxies `face-on' relative to their stellar angular momentum and binning the relevant quantities into a Cartesian grid with a pixel size of 750 parsecs (pc) on a side and following the methods of \citet{OrrEtAl2020SwirlsOfFIRE}.
We conduct two separate experiments aimed at finding an optimized parameterization of $\sigsfr$ using symbolic regression at two different  timescales for star formation - a measure of how quickly neutral gas is being converted to stars. One experiment finds parameterizations of star formation rate surface density averaged over 10 Myr, $\sigsfrTENMYR$, which would observationally correspond to tracers such as H$\alpha$, thus would be more sensitive to stochastic processes such as supernovae or various feedback processes. The other target variable to optimize an equation-based model for is that of star formation rate surface density averaged over 100 Myr, $\sigsfrHUNDREDMYR$, corresponding to observations traced by ultraviolet (UV) wavelengths or SED fitting. As the longer SF timescale smooths out stochastic processes and thus is relatively insensitive to them, we pursue these two independent experiments in order to explore and analyze any potential differences in functional form and/or the strength of the variables that are found to describe the two target variables respectively. From this point onward, the averaging timescale of the experiment will be denoted as $\tsfr$. Figure~\ref{fig:m12b_ZSFR} shows the spatially resolved $\tsfr$ = 10~Myr and $\tsfr$ = 100~Myr $\sigsfr$ maps of a single snapshot at $z$=0 of one of the simulated galaxies in our training dataset, m12b, as an example.

A summary of all the relevant features and the fiducial values for $L_*$ galaxies, which are bright galaxies comparable to the Milky Way, can be found in Table~\ref{tab:fiducial}. For specific details regarding the simulations and spatial mapping used in this study, please refer to the ``Simulations and Analysis Methods'' section of \citet{OrrEtAl2020SwirlsOfFIRE}. Our training dataset consists of all pixels from every galaxy snapshot across the sample in time where all the relevant features are co-spatially defined (i.e., all pixels where there is both recent star formation and gas). To leverage our ML techniques, these points are divided into a training set and a test set with a train-to-test split of $80\%$ to $20\%$.
The total number of usable pixels for $\tsfr$=10 Myr and $\tsfr$=100 Myr is $5.6\times10^4$ and $2\times10^5$ pixels respectively. Thus we acknowledge that the $\tsfr$=100 Myr results in more trainable pixels and therefore data points. We also acknowledge that our dataset is biased towards low-redshift pixels, with pixels exhibiting redshifts between 0 and 0.5 encompassing $64\%$ and $52\%$ for the $\tsfr$=10 Myr and $\tsfr$=100 Myr datasets, respectively.

 In this study, we are only concerned with local properties that are not influenced by whether the volume is contained within a disk or dispersion-supported environment.  \citet{Orr2018FIREKSlaw} found that for that for the galaxies within our dataset, between the redshifts of 0 to 6 there is no significant redshift evolution in the spatially resolved KS relation. We, therefore, decided not to include redshift as an input feature or break up the data into multiple redshift bins.

\begin{table*}[ht]
\caption{List of fiducial parameters.}
\centering
\begin{tabular}{lll|ll|l}
    \hline 
                            &                             &                                        &             &\\
    Parameter               & Units         & Description                                          & FIRE-2 & Local $L_*$ & Reference \\
                            &               &                                                      & range  & average     &           \\
    \hline 
                            &                             &                                        &         &             &\\
    $\Sigma_{\mathrm{gas}}$ & $\mathrm{M_{\odot}pc}^{-2}$ & Gas surface density                    & 10-510  & 10          & \citet{Wyder2009GALEXDwarfs, Kennicutt:1998aa} \\
                            &                             &                                        &         &             &\citet{Bigiel:2008aa, DaddiEtAl2010SFLaws}\\
    $\Omega_{\mathrm{dyn}}$ & Gyr$^{-1}$                  & Local orbital dynamical frequency      & 8-690   & 30          & \citet{KrumholzBurkhart2018GalacticDiscsModel} \\
    $\sigma_{\mathrm{gas}}$ & km s$^{-1}$                 & Gas velocity dispersion                & 6-310   & 10          & \citet{KrumholzBurkhart2018GalacticDiscsModel} \\
    $f_{\mathrm{gas}}$      &                             & Local gas fraction ($\Sigma_{\rm gas}/(\Sigma_{\rm gas}+\Sigma_\star)$) & 0-1     & 0.2         & \citet{LeroyEtAl2021} \\
    $\Sigma_{*}$            & $\mathrm{M_{\odot}pc}^{-2}$ & Stellar surface density                & 0-11000 & 100         & \citet{MeraEtAl1998}, \citet{SarkarJog2018}\\
    $\sigma_{\mathrm{*}}$   & km s$^{-1}$                 & Stellar velocity dispersion            & 20-200  & 45          & Physical arguments given fiducial \\
                            &                             &                                        &         &             & stellar disk scale height $h_*$=1.5 kpc\\
    $V_\mathrm{gas,\phi}$   & km s$^{-1}$                 & Velocity of gas in $\phi$ direction    & 0-360   & 240         & \citet{ReidDame2016}                            \\
    $V_\mathrm{gas,z}$      & km s$^{-1}$                 & Velocity of gas in $z$ direction       & 0-220   & 0           & Physical arguments\\
                            &                             &                                        &         &             &\\
    \hline 
\end{tabular} \label{tab:fiducial}
\end{table*}

\section{Results and Analysis for Equation Search in FIRE-2 simulations} \label{sec:results_and_analysis}

\subsection{Comparable models} \label{subsec:comparable_models}

\begin{table*}[ht]
\caption{ML and analytic models for SF for comparison with results of SR equation search pipeline}
\centering
\begin{tabular}{llll}
    \hline 
                                                                                                                                      &                                &         &           \\
    Model Equation                                                                                                                    & Model Reference                & Acronym & Model Category \\
                                                                                                                                      &                                &         &           \\
    \hline 
                                                                                                                                      &                                &         &           \\
    $\sigsfr := \mathbf{\Omega}(\Sigma_{\mathrm{gas}},~\Omega_{\mathrm{dyn}},~\sigma_{\mathrm{gas}},~f_g,~\Sigma_{*},~\sigma_{\mathrm{*}},~V_\mathrm{gas,\phi},~V_\mathrm{gas,z} )$ &\citet{Chen2016XGBoost}        & XGBoost & ML regression decision tree \\
    $~~~~~~~ \mathrm{for~trained~model~} \mathbf{\Omega}$                                                                             &                                &         &           \\
                                                                                                                                      &                                &         &           \\
    $\sigsfr \propto \Sigma_{\mathrm{gas}} ^{1.4}$                                                                                    &\citet{Kennicutt:1998aa}        & K98     & Empirical \\
                                                                                                                                      &                                &         &           \\
    $\sigsfr =  \epsilon_{\mathrm{ff}}\frac{\siggas}{\tff},$                                                                          &\citet{KrumholzDekelMcKee2012}  & KDM12   & Bottom-up $\&$ empirical \\
    $~~~~~~~ \mathrm{where}~t_{\mathrm{ff}} = \sqrt{\frac{3\pi}{32G\rho}}$                                                            &                                &         &           \\
                                                                                                                                      &                                &         &           \\
    $\sigsfr =  \epsilon_{\mathrm{ff}} \frac{\siggas}{\tff}\Big(1+b^2\mathcal{M}^2\frac{\beta}{\beta + 1}\Big)$                       &\citet{SalimEtAl2015}                 & SFK15   & Bottom-up \\
                                                                                                                                      &                                &         &           \\
    $\sigsfr =  \epsilon_{\mathrm{ff}} \frac{1}{2}\frac{\siggas}{\tff}\Big(1+b^2\mathcal{M}^2\frac{\beta}{\beta + 1}\Big) \Big[ 1+ \mathrm{erf}\Big(\frac{\sigma_s^2 - s_{\mathrm{crit}}}{\sqrt{2 \sigma_s^2}} \Big)\Big]$, &\citet{Salim2020HCG} & SAF20   & Bottom-up \\
    $~~~~~~~\mathrm{where}~ s_{\mathrm{crit}} = \ln \Big[\Big(\frac{\pi^2}{5}\Big)\phi^2_x \alpha_{\mathrm{vir}}\mathcal{M}\Big]$     &                                &         &            \\
                                                                                                                                      &                                &         &            \\
    $\sigsfr = \frac{\sqrt{3}}{2} \frac{\Sigma_{\mathrm{gas}}\Omega_{\mathrm{dyn}}\sigma_{\mathrm{gas,z}}}{(P_*/m_*)}$                &\citet{FaucherGiguere13}        & FG13    & Top-down   \\
                                                                                                                                      & (identical scaling to          &         &\\
                                                                                                                                      & \citet{OstrikerKim2022},       &         &\\
                                                                                                                                      & $\&$ in limiting case where $f_\mathrm{gas}$=1,&         &\\
                                                                                                                                      & to \citet{OstrikerShetty2011})                  &         &\\
                                                                                                                                      &                                &         &           \\
    \hline 
\end{tabular} \label{tab:sf_laws_analytic}
\end{table*}


\begin{figure*}
    \includegraphics[width=0.5\linewidth]{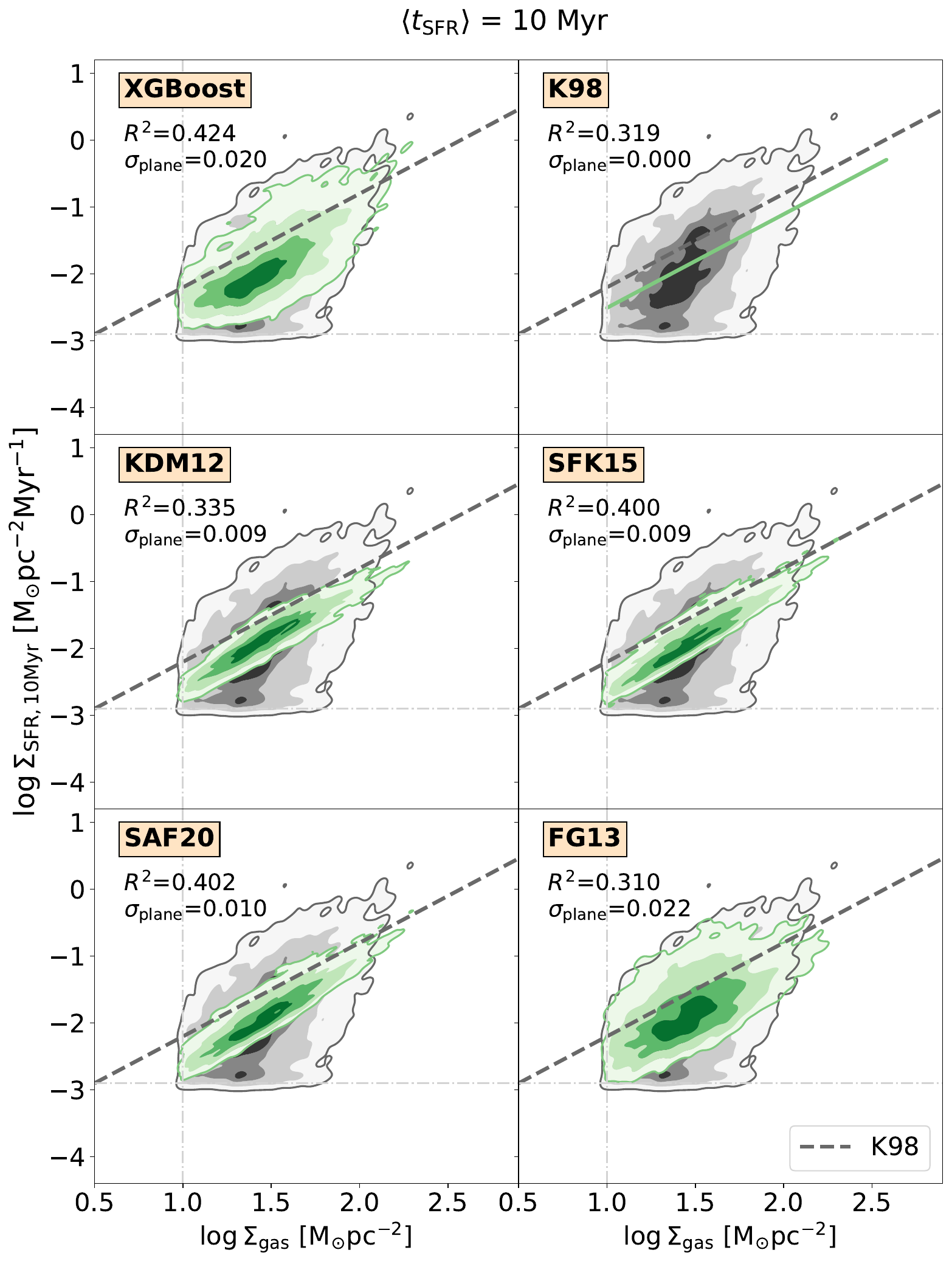}
    \includegraphics[width=0.5\linewidth]{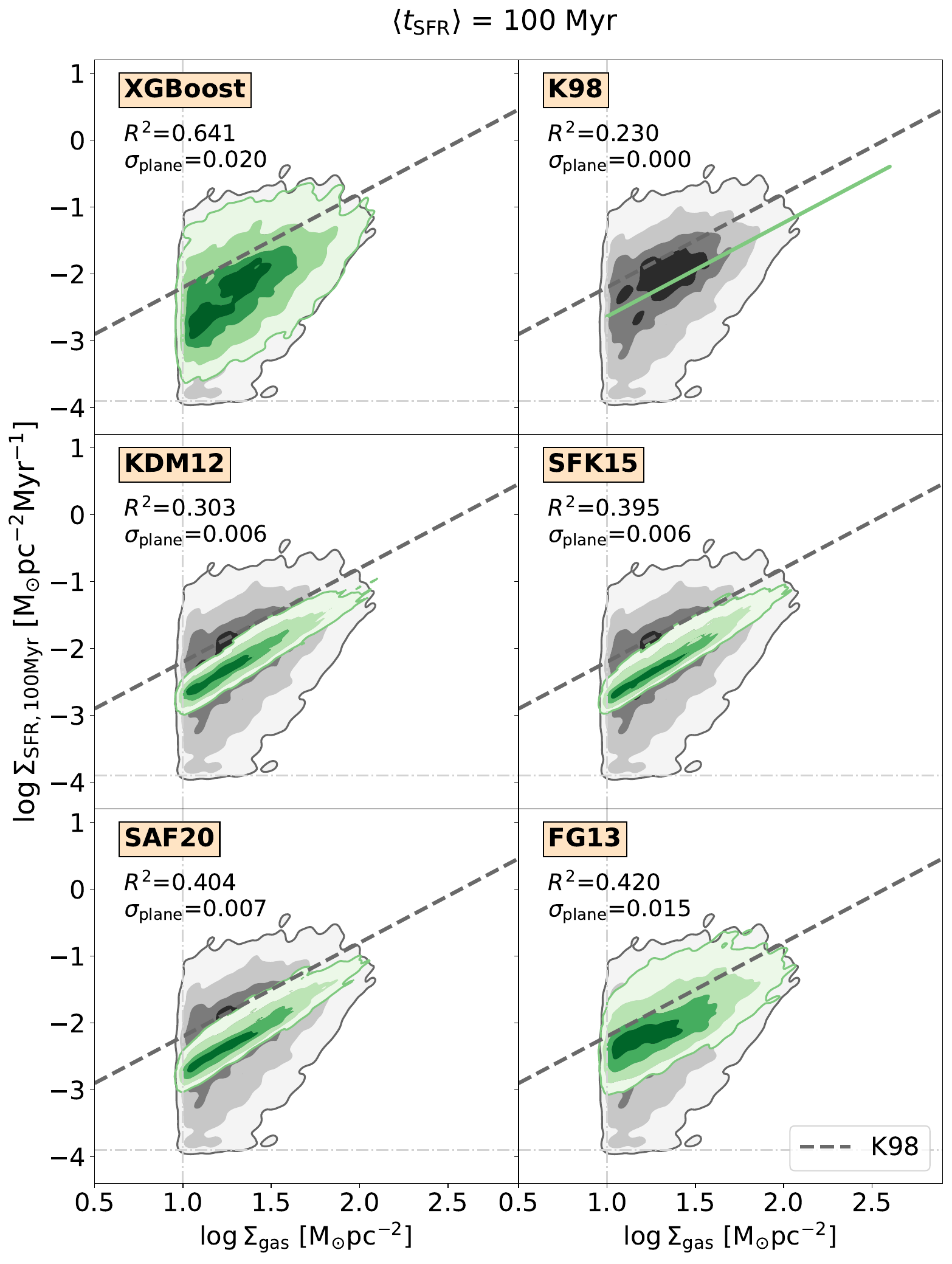}
    \caption{Distribution of points across the KS plane of the FIRE data (grey contours) predicted by the respective models labeled in the top-left orange boxes for the $\tsfr$=10 Myr (\textit{left subfigure}) and $\tsfr$=100 Myr (\textit{right subfigure}) experiments. For all contours (grey and green), the regions represent the $95\%$, $80\%$, $50\%$ and $20\%$ data inclusion region in the plane (from lightest to darkest colored, outside-in). The dashed, dark grey line shows the original K98 relation. The vertical line shows the $\siggas$ cut of $\siggas$=10 $\mathrm{M_{\odot}pc^{-2}}$, points below which we do not consider in this study, and the horizontal line shows the numerical SF floor in the simulations set by $\Sigma_{\mathrm{SFR,~floor}} = m_\mathrm{b}/(A\cdot \langle t_{\mathrm{SFR}}\rangle)$, where $m_\mathrm{b}$ is the minimum baryonic mass of 7100 $\mathrm{M_{\odot}}$, $A$ is the simulation's cartesian projection pixel size of $(750~\mathrm{pc})^2$ in our input maps and $\langle t_{\mathrm{SFR}} \rangle$ is the time over which the SF is averaged over, in this case either 10 or 100 Myr.} \label{fig:SFR_comparable_models}
\end{figure*}

\begin{table*}[ht]
\caption{Table of the goodness-of-fit, $R^2$, value and the standard deviation of the line of best fit through the points predicted by the model, $\sigma_{\mathrm{plane}}$, for each of the analytic models considered in this study. Statistically best values are shown in bold font.}
\centering
\begin{tabular}{l|l|ll|lllll}
    \hline
                          &                               &            &         &     &       &       &       &      \\
                          &                               &FIRE-2 test & XGBoost & K98 & KDM12 & SFK15 & SAF20 & FG13 \\
                          &                               &            &         &     &       &       &       &      \\
    \hline
                          &                               &            &       &       &       &       &      &       \\
    10 MYR SFR TIMESCALE  & $R^2$                         & -          & 0.424 & 0.319 & 0.335 & 0.400 & \textbf{0.402}  & 0.310 \\
                          & $\sigma_{\mathrm{plane}}$  & 0.032      & 0.020 & 0.0   & 0.009 & 0.009 & 0.010  & \textbf{0.022} \\
                          &                               &            &       &       &       &       &      &       \\

    \hline
                          &                               &            &       &       &       &       &      &       \\
    100 MYR SFR TIMESCALE & $R^2$                         & -          & 0.641 & 0.230 & 0.303 & 0.395 & 0.404  & \textbf{0.420} \\
                          & $\sigma_{\mathrm{plane}}$  & 0.026      & 0.020 & 0.0   & 0.006 & 0.006 & 0.007  & \textbf{0.015} \\
                          &                               &            &       &       &       &       &      &       \\
    \hline
\end{tabular}\label{tab:analytic_model_metrics}
\end{table*}

In this study, we compare the equations found by \textsc{PySR} to six models as summarized in Table~\ref{tab:sf_laws_analytic}. The comparable models include one ML regression model, XGBoost, which was retrained on the final 8 variables described in Section~\ref{subsubsec:data_prep}, the SHAP analysis of which is shown in Appendix~\ref{appendix:shap}. The theoretical analytic models being considered are the classic empirical K98 scaling relation (with the normalization of the power law refit such that the offset is optimized to the FIRE-2 test dataset), three analytic SF models that could be categorized as "bottom-up" (the efficiencies of which have all been fit for the FIRE-2 test dataset) and finally, a "top-down" categorized SF model. The distribution of $\sigsfr$ in each model is attained by inputting the relevant variables from our FIRE-2 simulation test data set. We only analyze behavior when applied to the test data set in order to have a fair comparison between the analytic models and the ML-based models that can only be validated on a test set.

We utilize two statistical metrics to assess how closely each model replicates the FIRE-2 simulations: (1) the goodness-of-fit, $R^2$, which is the coefficient of determination and has a range of 0–1 ($R^2$ = 1 indicates that a model that perfectly fits the data), and (2) the standard deviation of the line of best fit through the points predicted by the model, $\sigma_{\mathrm{plane}}$, a value as close to that of the native FIRE-2 simulation of which indicates a model can capture more information than just the mean trend. The latter metric tracks the dispersion across the KS plane, which arises due to a combination of physical processes in the simulation. These values are tabulated in Table~\ref{tab:analytic_model_metrics}, where the model which achieves the statistically best values as determined by these two metrics, for each of our separate experiments, are highlighted in boldface. These values are also recorded on the panels of Figure~\ref{fig:SFR_comparable_models}, which show how these models behave on the KS plane when the corresponding input variables from the FIRE-2 test dataset are used to predict $\sigsfr$ for each respective model (\textit{green contours}), compared against the FIRE-2 test dataset (\textit{grey contours}) for both the $\tsfr$=10 Myr (\textit{left subfigure}) and $\tsfr$=100 Myr (\textit{right subfigure}) experiments.

All models investigated here perform better than the traditional K98 relation on both metrics. The best-performing analytic models based on $R^2$ calculations are the SAF20 multi-freefall "bottom-up" model and the FG13 "top-down" models for the $\tsfr$=10 Myr and $\tsfr$=100 Myr experiments, respectively. XGBoost represents the degree to which a model can \textit{possibly} best map a set of input variables to a target variable for a specific dataset, as it is unbounded by the need to be expressed by an equation. This is demonstrated quantitatively for both the $\tsfr$=10 Myr and $\tsfr$=100 Myr SFR test datasets, in which it achieves the highest $R^2$ compared to all the models explored in this study. However, as it is difficult to interpret the mapping between input and target variables with this ML method, this model is challenging (if not impossible) to physically interpret. In the $\tsfr$=10 Myr SFR experiment, the one "top-down" model, FG13, achieves the lowest $R^2$ value but best recovers the dispersion across the KS plane, even achieving a $\sigma_{\mathrm{plane}}$ value closer to the FIRE-2 dataset on the KS plane than the XGBoost model. The "bottom-up" models, KDM12, SFK15, and SAF20, on the other hand, achieve increasing $R^2$ values, respectively, reaching close to the XGBoost $R^2$ upper limit, but do not adequately capture the dispersion across the KS plane, as visually evident in Figure~\ref{fig:SFR_comparable_models}, where the green contours which represent the distribution from the $\sigsfr$ values predicted from these models clearly appear much narrower than the native FIRE-2 distribution shown in the background grey contours.

In the $\tsfr$=100 Myr SFR case, there is a larger discrepancy between the $R^2$ values of all the analytic/empirical models and the upper limit set by XGBoost. However, the "top-down" model, FG13, achieves the highest $R^2$ and dispersion across the KS plane amongst the analytic models considered in this study. On this longer timescale, the "bottom-up models" are similarly deficient in capturing the KS plane dispersion, mirroring the behavior of these models applied to the $\tsfr$=10 Myr experiment.

\subsection{Loss curves $\&$ Performance of equations found by \textsc{PySR}} \label{subsec:eqn_performance}


\begin{figure*}
    \includegraphics[width=\linewidth]{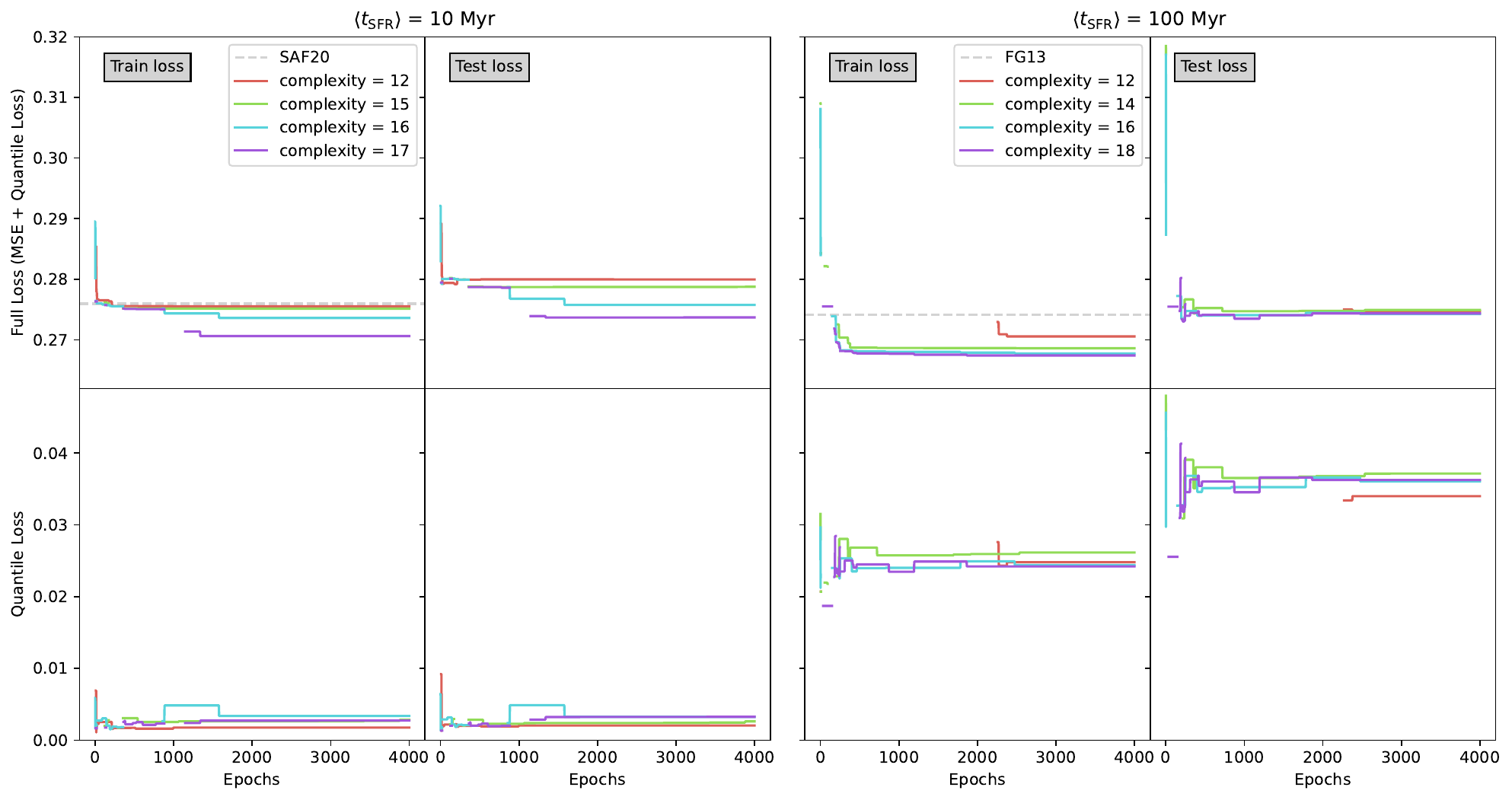}
    \caption{Evolution of the total loss values as defined by Equation~\ref{eq:LOSS_EQN} (\textit{top rows}) and the quantile loss values as defined by Equation~\ref{eq:quantile_loss} (\textit{bottom rows}) by training epoch for the 10 Myr averaged SFR target variable experiment (left subfigure) and the 100 Myr averaged SFR target variable experiment (right subfigure). The colored lines indicate the losses as defined by Equation~\ref{eq:LOSS_EQN} for the training (left columns within subfigures) and test (right columns within subfigures) data samples as a function of the number of training epochs for the top equations as selected by the method described in Section~\ref{sec:methods}. Note that whilst some equations exhibit the same complexity and therefore represented by the same line color between the $\tsfr$ = 10 Myr SFR and $\tsfr$ = 100 Myr SFR experiments, they are indeed representing different equations unique to the experiment of that specific SFR timescale. The grey dotted line shown in the panels of the training loss curves indicates the lowest scoring loss among the analytic models considered in this study. This is the "baseline" loss for which the top selected equations must exhibit a lower loss than.} \label{fig:loss_curves}
\end{figure*}


\begin{table*}[ht]
\caption{Top 4 equations that \textsc{PySR} found after training for 4000 epochs. The row highlighted in gray represents the best equations found as per our statistical analysis.}
\centering
\begin{tabular}{l|llll}
    \hline
                   &                  &       &       &         \\
    Experiment     & Model complexity & $R^2$ & Score & Equation\\
                   &                  &       &       &         \\
    \hline
    \\
    $\tsfr=10$ Myr         & 12               & 0.397 & 0.0113 & $\log \Sigma_{\mathrm{gas}}^{1.52} + \left(e^{\log f_{\mathrm{gas}}}\right)^{-0.4} - 4.98$ \\
                           & 15               & 0.401 & 0.0014 & $\left(e^{\log \Sigma_{\mathrm{gas}}} + \left(e^{\log \Sigma_{*}}\right)^{0.49}\right)^{0.64} - 5.33$ \\
                           & 16               & 0.409 & 0.0056 & $\log \Sigma_{\mathrm{gas}}^{1.26} + \left(e^{\log \Sigma_{*}}\right)^{0.23} + \frac{\log{\left(\log \sigma_{\mathrm{gas},z} \right)}}{\log{\left(10 \right)}} - 5.19$ \\
    \rowcolor{lightgray} &  17 & 0.413 & 0.0110 & $\log \Sigma_{\mathrm{gas}}^{1.29} + \left(e^{\left(\log \sigma_{\mathrm{gas},z} + \log \Sigma_{*}\right)^{1.25}}\right)^{0.12} - 5.21$\\
    \\
    \hline
    \\
    $\tsfr=100$ Myr        & 12                      & 0.550 & 0.0343 & $\log \Sigma_{\mathrm{gas}} + \left(\log \sigma_{\mathrm{gas},z} + \log \Sigma_{*}\right)^{0.86} - 6.12$ \\
     \rowcolor{lightgray}  & 14 & 0.555 & 0.0036 & $\log \Sigma_{\mathrm{gas}} + \left(\log \sigma_{\mathrm{gas},z} + \log \Sigma_{*}\right)^{0.85} - 6.08$\\
                           & 16                      & 0.555 & 0.0016 & $\log \Sigma_{\mathrm{gas}} + \left(\log \sigma_{\mathrm{gas},z}^{0.9} + \log \Sigma_{*}\right)^{0.86} - 6.06$\\
                           & 18                      & 0.555 & 0.0006 & $\log \Sigma_{\mathrm{gas}}^{0.96} + \left(\log \sigma_{\mathrm{gas},z}^{0.91} + \log \Sigma_{*}\right)^{0.86} - 6.06$\\
    \\
    \hline
\end{tabular}\label{tab:final_eqns}
\end{table*}

\begin{figure*}
    \includegraphics[width=\linewidth]{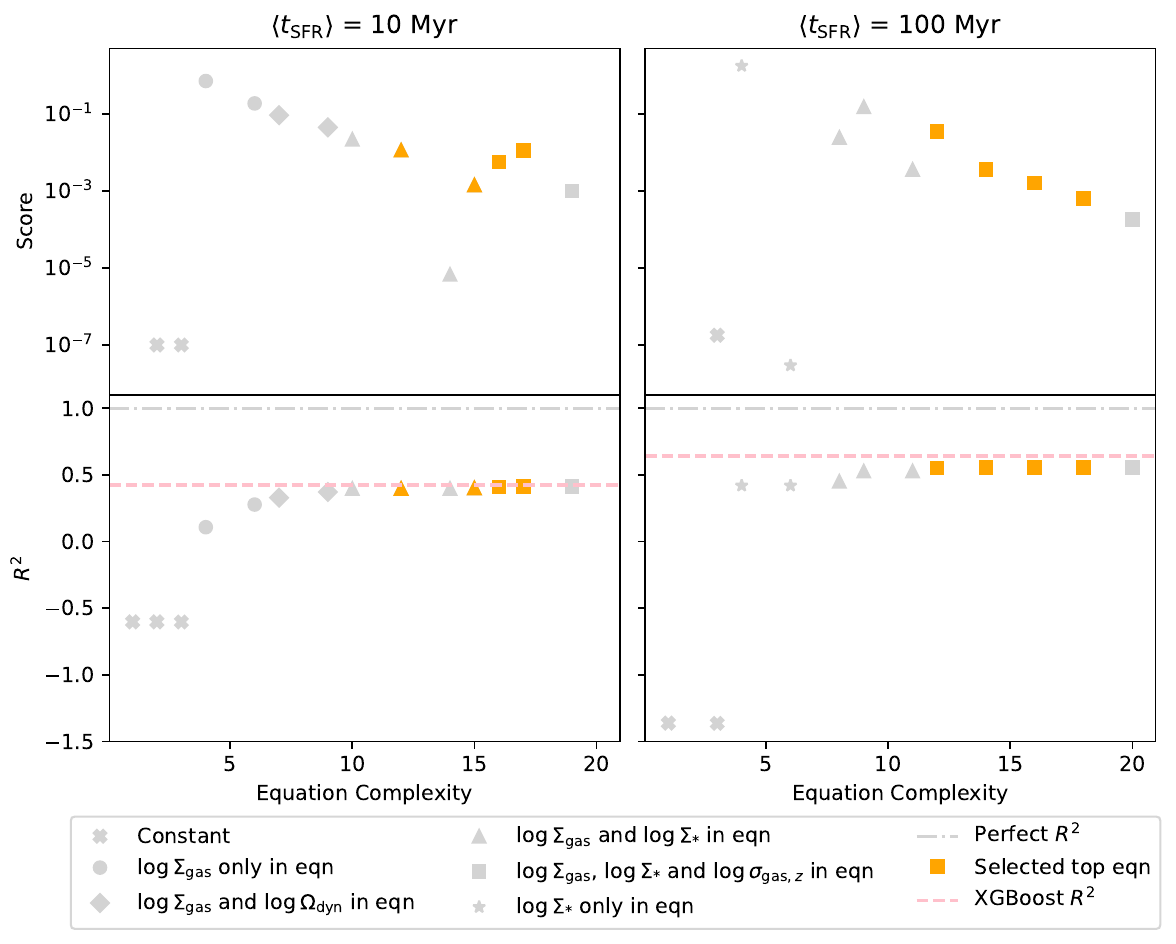}
    \caption{Points indicating the equation "score" as originally defined by \citet{Cranmer2023PySR} (top row) and the $R^2$ (bottom row) of all equations found during the 4000 epoch training run for the two experiments when SF is averaged over $\tsfr$ = 10 Myr (\textit{left}) and when averaged over $\tsfr$ = 100 Myr (\textit{right}). In all panels the orange points indicate the models that satisfy the criteria to count as a "top equation", as highlighted in Section~\ref{sec:methods}. In the bottom panel, the grey dash-dot lines are indicative of a perfect $R^2$ value of 1, which occurs when the model and the data are in exact agreement. The dashed pink lines are indicative of the $R^2$ value that the XGBoost model achieves when trained on the same dataset, and represents the upper limit to an $R^2$ value that equation-based models can achieve. The shapes of the points showcase the combination of variables that feature in each model. Crosses indicate models which are described by just one constant value, circle points indicate that only $\siggas$ is present in the model, star-shaped points indicate that only $\sigstars$ are present in the model, diamond-shaped points indicate that $\siggas$ and $\Omega_{\mathrm{dyn}}$ are present in the model, triangle points designate models featuring both $\siggas$ and $\sigstars$, and square points specify models constructed via all three of the features $\siggas$, $\sigstars$ and $\gasvdisp$. Note that there exist 1 and 2 models in the $\tsfr$ = 10 Myr and $\tsfr$ = 100 Myr experiments, respectively, that yielded a score of 0, thus do not appear in the top panel of Figure~\ref{fig:train_metrics} which features a logarithmic scale.}
    \label{fig:train_metrics}
\end{figure*}
\begin{figure*}
    \includegraphics[width=\linewidth]{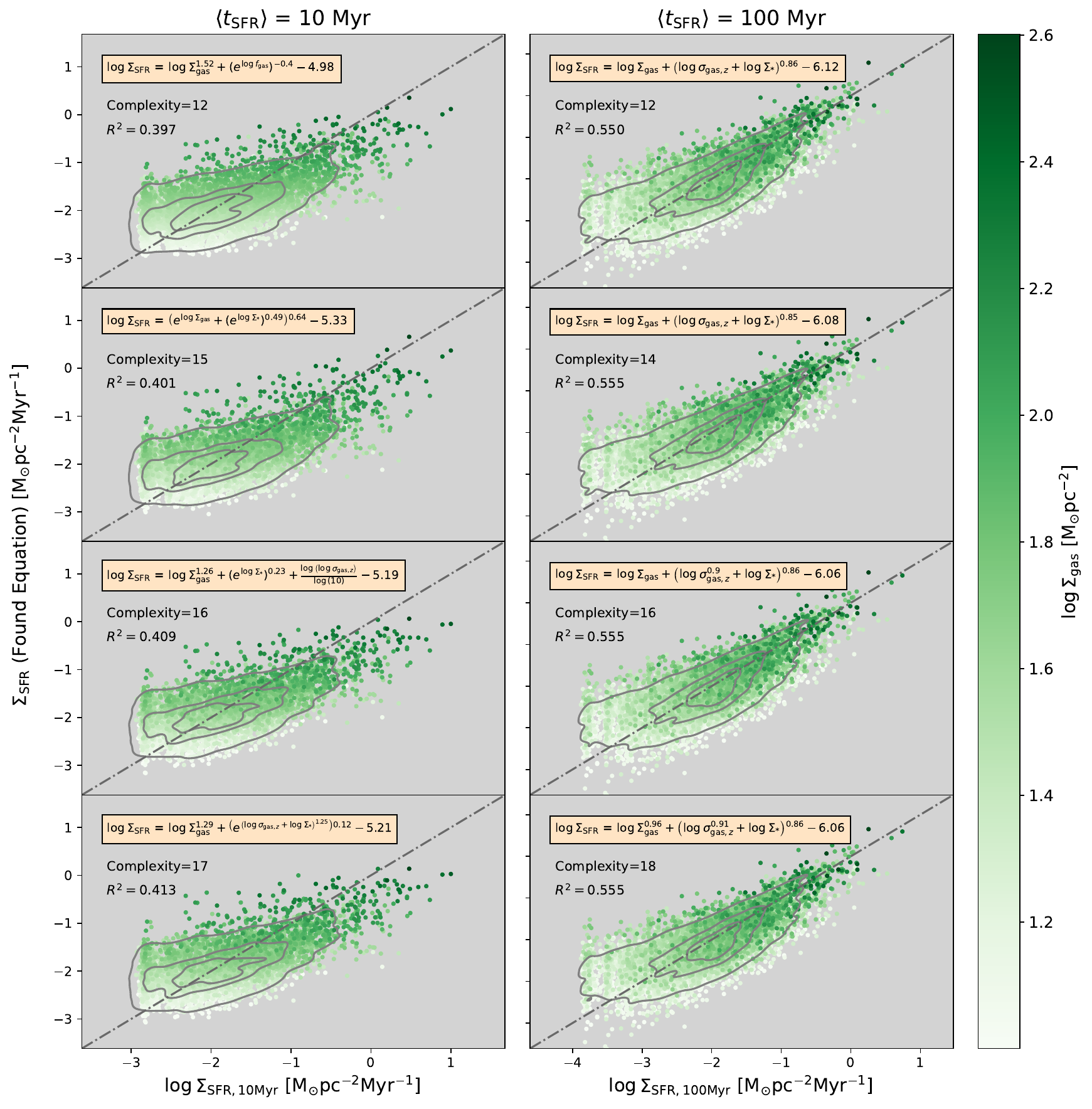}
    \caption{The predicted $\Sigma_{\mathrm{SFR}}$ of the top four models found by PySR in comparison to the native $\Sigma_{\mathrm{SFR}}$ of the FIRE-2 test sample, with the colour indicating the value of $\Sigma_{\mathrm{gas}}$ at that point for the $\tsfr$ = 10 Myr \textit{(left column)} and 100 Myr \textit{(right column)} experiments. The overlaid contours represent, from the outer to innermost contours, where $90\%$, $50\%$ and $20\%$ of the probability mass of the data in this plane lies. The found equation itself is shown in the orange box in the top left hand corner of each panel. The one-to-one line is shown as the gray dash-dot line. The complexity of the equation and the $R^2$ value of the model is written directly beneath the equation. A model that model the predicts the data perfectly would exhibit an $R^2$ of 1, and all points would lie on the one-to-one line.}
    \label{fig:SFR_one_to_one}
\end{figure*}

Post-training, the loss as defined by Equation~\ref{eq:LOSS_EQN} is calculated for each of the analytic models considered in this study and the model (excluding XGBoost) that achieves the lowest scoring loss (and therefore highest $R^2$ values) for each of the $\tsfr$ = 10 Myr and $\tsfr$ =100 Myr experiments' respective test data sets are taken as the "baseline" loss. For the $\tsfr$ = 10 Myr experiment, the best performing analytic model is the "bottom-up" \citet{Salim2020HCG} multi-freefall model and for the $\tsfr$ = 100 Myr experiment it is the "top-down" \citet{FaucherGiguere13} model. The \textsc{PySR}-found equations that we select as our "top selected equations" are the top four equations that exhibit final training losses that are less than this baseline loss, ranked in order of the "score" as applied by \citet{Cranmer2023PySR}.

The evolution of the train and test losses as a function of the training epoch is presented in Figure~\ref{fig:loss_curves}, with the full losses shown on the top and the isolated quantile losses shown on the bottom rows, respectively. The ratio between test loss and train loss is always consistently around 1, indicating that the models are not overfitting to the training data. We see that at equal weighting of mean square error loss and quantile loss to the total loss, the quantile loss only contributes $10\sim 15\%$ to the total loss. This indicates that the loss is thus fitting predominantly to the mean and not accounting as much for the variance of the data. The final resultant equations attained by the end of this training are tabulated in Table~\ref{tab:final_eqns}.

In Figure~\ref{fig:train_metrics} we present a snapshot of \textit{all} the models found by \textsc{PySR} for both experiments by showing how the "score" and $R^2$ metric of the models change with increasing model complexity (not just the selected top ones, which are highlighted in orange). These figures help defend our justification for our method in choosing the top equations. We see that if we had adhered to assessing the equations by the sole rank in their "scores", as expected, we would have yielded very simple equations that described SF only by one variable. Whilst this shows that these equations have recovered KS-law-like scaling relations, which is encouraging as it is unsurprising, they do not inform us of any additional nuances of star formation as more complex analytic models do. However, consideration of these models found at lower complexities can provide insight into which variables contain the most information density when describing the target variable. It is evident as in the equation complexity versus $R^2$ relations shown in the bottom row of Figure~\ref{fig:train_metrics}, that the lower complexity models indicate that if $\sigsfrTENMYR$ were to be traced by just one variable, $\siggas$ would hold the most information density, while $\sigsfrHUNDREDMYR$ is dominated by information contained in $\sigstars$. This indicates that gas is the most influential factor for determining star formation in datasets that reflect more stochastic processes. Whereas, in the $\tsfr$=100 Myr case the timescale averages over several cloud lifetimes and therefore reflects a situation closer to equilibrium. This appears to be best described by the stellar surface density, which is closely related to the potential of the disk. This finding is consistent with the SHAP analysis as shown in Figure~\ref{fig:shap_summary_plot} in Appendix~\ref{appendix:shap}, which shows the prediction that the variables $\siggas$ and $\sigstars$ should best correlate with the target variables $\sigsfrTENMYR$ and $\sigsfrHUNDREDMYR$, respectively.

We see that the $R^2$ values between the model and equations found by \textsc{PySR} generally increases with complexity and converge to a value close to that of the XGBoost's. Given that the $R^2$ value indicates the proportion of the variance in the target variable that can be explained by the dependent features, and that the $R^2$ value of the XGBoost model represents the highest possible variance that can be captured, because the top equations chosen by our algorithm exhibits $R^2$ of or greater than 0.397 and 0.550 for the $\tsfr$ = 10 Myr and 100 Myr experiments respectively, this conveys that our found models have captured at least 93$\%$ and 85$\%$ respectively of the maximum possible variance that can be captured. This indicates that the found models reflect close to the most information that is possible to contain within a model given the input variables tested, which gives confidence that these are the best possible models found.

However, we see that there is a diminishing return in an increase of $R^2$ value with increasing complexity. This is evident in the bottom rows of Figure~\ref{fig:train_metrics} where we see the orange points denoting the selected top equations to approach XGBoost's $R^2$ value at higher complexities, but the increase in $R^2$ between the lowest and highest complexity amongst the top found models are minuscule. For both datasets, it is clear that $\siggas$ and $\sigstars$ take dominant influence over $\sigsfr$, with the addition of $\gasvdisp$ also contributing further lesser-order information density at higher-complexity models, a finding that is once again exactly consistent with the SHAP analysis of these datasets.

We present the actual top equations in full written form in Table~\ref{tab:final_eqns}, which are reiterated for ease of reference in the panels of Figures~\ref{fig:SFR_one_to_one}--\ref{fig:SFR100Myr_KSbehaviour}. The variables from the FIRE-2 test data set that appear in each model are input into each equation and the resulting $\sigsfr$ values are evaluated against the native values of $\sigsfr$ from the FIRE-2 test data set, therefore allowing us to assess the degree to which each model agrees to the native FIRE-2 test dataset via the $R^2$ metric. For both $\tsfr$=10 Myr and 100 Myr, we observe that some combination of $\siggas$, $\sigma_{\mathrm{gas,~z}}$ and $\sigstars$ feature prominently in all equations found by PySR (as $f_\mathrm{gas}$ is a derivative of $\siggas$ and $\sigstars$). This finding is consistent with the SHAP values' evaluation of the importance of these features in influencing SF, as highlighted in Figure~\ref{fig:shap_summary_plot} and Section~\ref{subsec:FIREsims}. It is encouraging that $\siggas$ features very prominently in all found equations, as this is consistent with many known theories for star formation such as \citet{Kennicutt:1998aa} and all the other analytic models tested in this study as tabulated in Table~\ref{tab:sf_laws_analytic}.

\subsection{Feature space planes} \label{subsec:feature_space_planes}
\begin{figure*}
    \includegraphics[width=\linewidth]{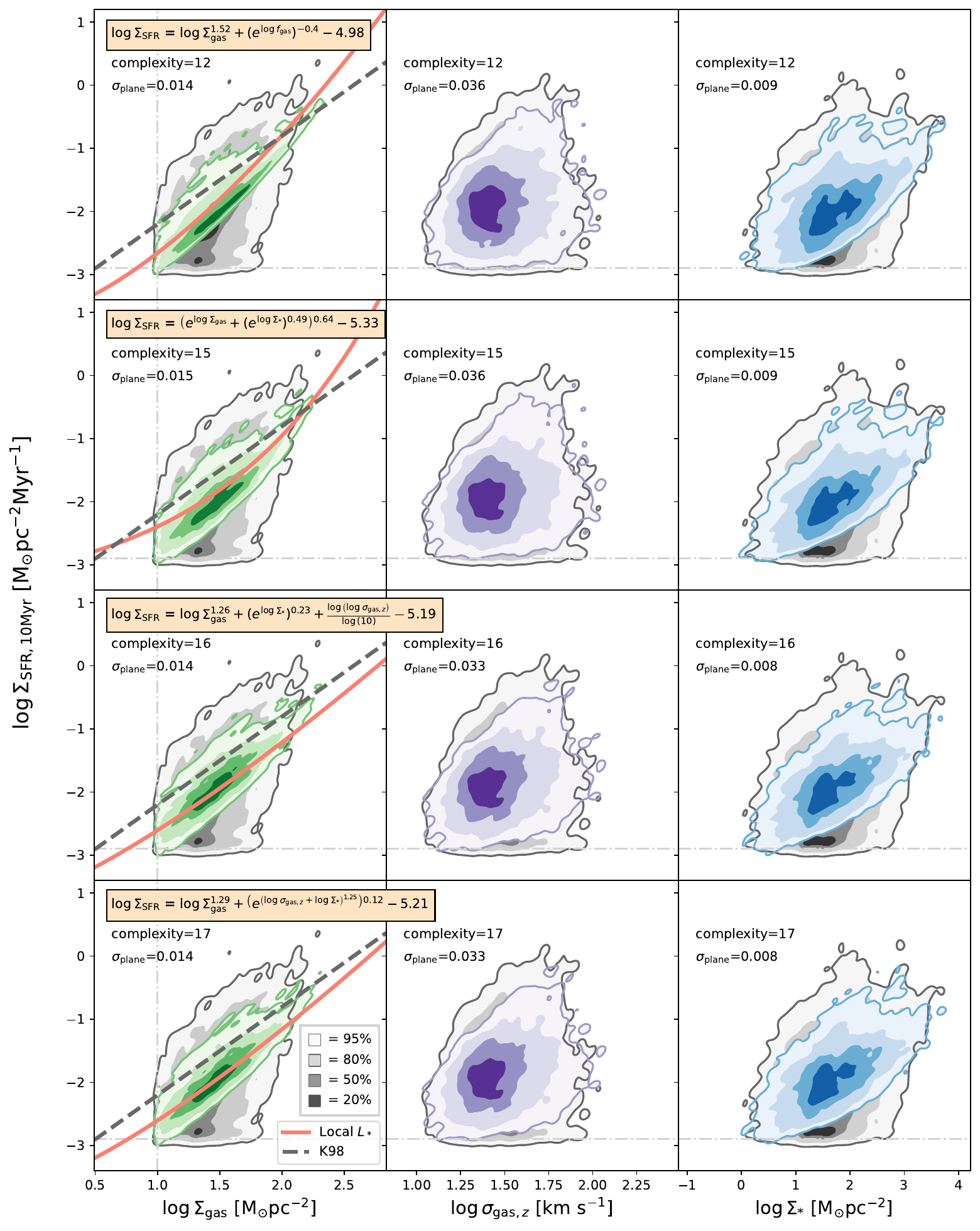}
    \caption{ Distribution of points across the KS (logarithmic $\siggas$-$\sigsfr$) plane, the logarithmic $\sigma_{\mathrm{gas,z}}-\sigsfr$ plane and the $\sigstars-\sigsfr$ planes of the native FIRE data (grey contours) and $\Sigma_{\mathrm{SFR}}$ attained for the respective equations found by PySR (colored contours; green, purple and blue for $\siggas$, $\gasvdisp$ and $\sigstars$ respectively). For both categories of contours, the outer (lightest colored) to innermost (darkest coloured) regions represent where $95\%$, $80\%$, $50\%$ and $20\%$ of the probability mass of the data in this plane lies. The dashed line shows the original K98 relation. The pink line shows how the PySR-found equation would behave on this plane when the features other than $\Sigma_{\mathrm{gas}}$ are set to the fiducial values for Local Spiral galaxies as tabulated in Table~\ref{tab:fiducial}.}
    \label{fig:SFR10Myr_KSbehaviour}
\end{figure*}

\begin{figure*}
    \includegraphics[width=\linewidth]{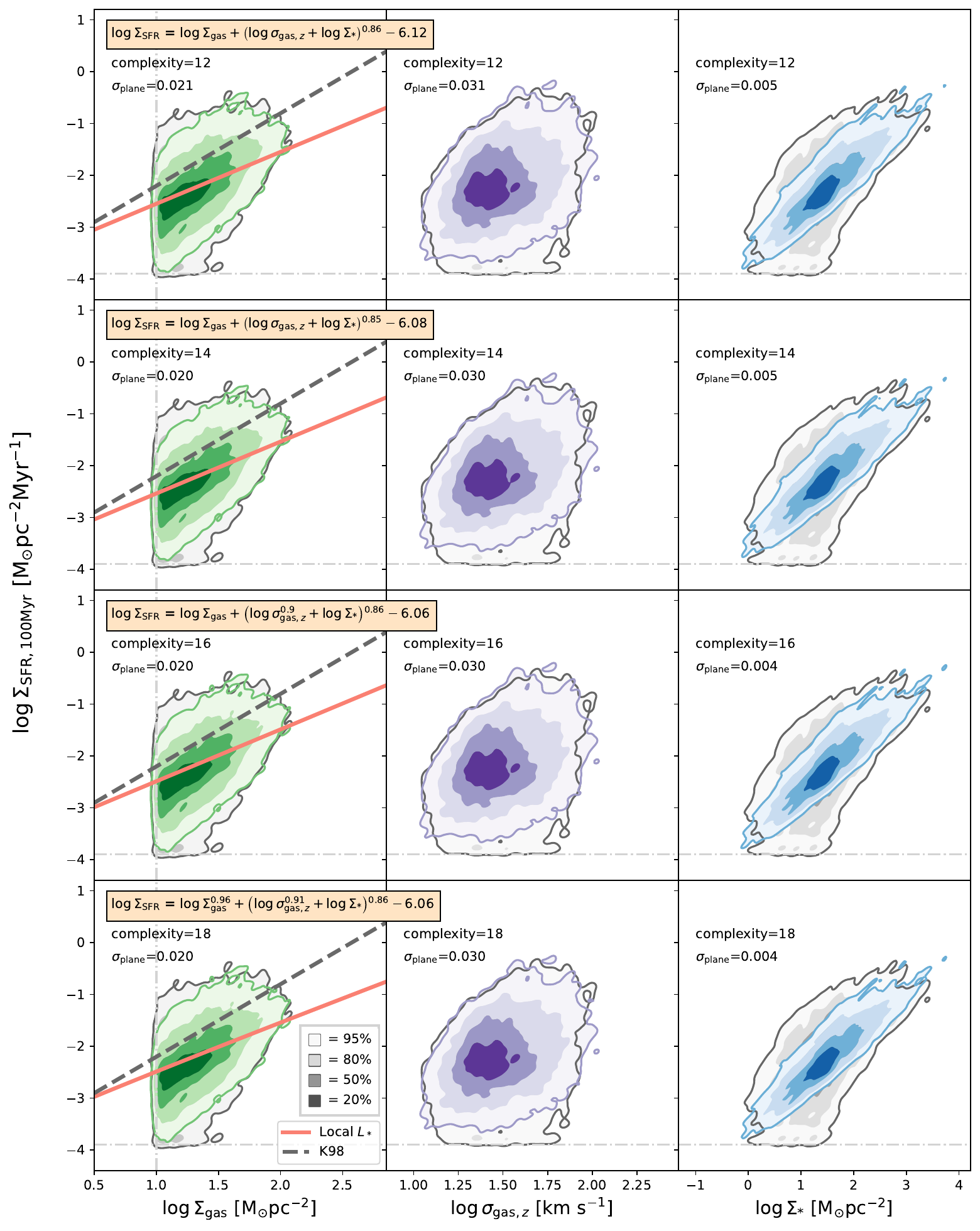}
    \caption{Same as in Figure~\ref{fig:SFR10Myr_KSbehaviour}, but showing the results for the $\langle t_{\mathrm{SFR}}\rangle$ = 100 Myr experiment. Note that only five equations met the criteria to be considered a top equation.}
    \label{fig:SFR100Myr_KSbehaviour}
\end{figure*}

To explore the independent contributions of each variable that appears in the models to the target variable and analyze the behavior of the equations found by \textsc{PySR} in a space that is comparable to observations, in Figures~\ref{fig:SFR10Myr_KSbehaviour} and~\ref{fig:SFR100Myr_KSbehaviour}, we present the 2D distribution of points across the planes of features $\siggas$, $\gasvdisp$ and $\sigstars$ against the target variable $\sigsfr$, for $\tsfr$ = 10 Myr and $\tsfr$ = 100 Myr experiments respectively. These planes will be referred to as "feature space planes" for each contributing variable. As in Figure~\ref{fig:SFR_comparable_models}, we show how the found models behave on the feature space planes when the corresponding input variables from the FIRE-2 test dataset are used to predict $\sigsfr$ for each respective model, compared against the FIRE-2 test dataset. It is desirable that a model exhibits a distribution across the feature space planes that mimics the dispersion in these planes as the native FIRE-2 data, which represents the degree of stochasticity within the data or hidden correlations.

We first consider the feature space planes of $\siggas$ (also known as the KS plane) and that of $\sigstars$. Whilst the influence of $\siggas$ on $\sigsfr$ is abundantly apparent and expected, concentrating on the left-most and right-most columns of Figures~\ref{fig:SFR10Myr_KSbehaviour} and~\ref{fig:SFR100Myr_KSbehaviour}, corresponding to the feature-to-target variable spaces of $\siggas$ and $\sigstars$, we see that both show that the found models are finding rough scaling relations between the target variable and the feature. We postulate that due to the similar radial profiles of $\siggas$ and $\sigstars$ that feature high densities towards the center of galaxies and decrease exponentially with distance, these two features are highly coupled to one another.

However, when considering the $\sigsfr$ distributions predicted by the top \textsc{PySR} found equations, we observe that the variables that demonstrate the most influence towards the target variable also display the weakest ability to capture the variance of the original data in their feature space planes. This indicates that these models are predicting a slightly stronger correlation between the most influential input variable and the target variable than is actually present in the original data. For $\tsfr$=10 Myr, as mentioned above, we observe that the $R^2$ of all the models found converge towards that of XGBoost, indicating that the best possible models for this given dataset have been found, yet they perform poorly at recreating the intrinsic scatter of the FIRE-2 test data in particularly the KS and, to a lesser degree, the $\sigstars$ feature space planes. This presents as a narrow dispersion and quantitatively as a value for the variance of the Gaussian fitted to the residuals of the best linear fit to the points in the feature space planes, $\sigma_{\mathrm{plane}}$, that is lower than that of the native FIRE-2 data.
On the other hand, the $\tsfr$=100 Myr experiment yields models that exhibit $R^2$ values that achieve convergence at a value of greater discrepancy to that of the XGBoost model but are overall greater values than that of the $\tsfr$=10 Myr experiment. We note that the distribution across the KS plane recovers that of the native FIRE-2 data much closer than the models found for the $\tsfr$ = 10 Myr does, but exhibits narrower distributions and lower $\sigma_{\mathrm{plane}}$ for $\sigstars$, which is the variable that best describes the $\sigsfrHUNDREDMYR$ dataset given it to be described by just one variable. It is therefore apparent that the more influence a variable has over the target variable, the more of a tendency that our \textsc{PySR}-based pipeline has to overestimate the target variable's dependency on it, thereby poorly recovering the distribution of the dataset in its feature space plane. This indicates that despite the exhaustive search for a predictive model for SF, with the symbolic regression method utilized in this study (including our choice of loss function), it is not possible to accurately fit a function that approaches both the maximum possible $R^2$ \textit{and} dispersion across the feature-space planes.

\begin{figure*}
    \centering
    \includegraphics[width=0.48\textwidth]{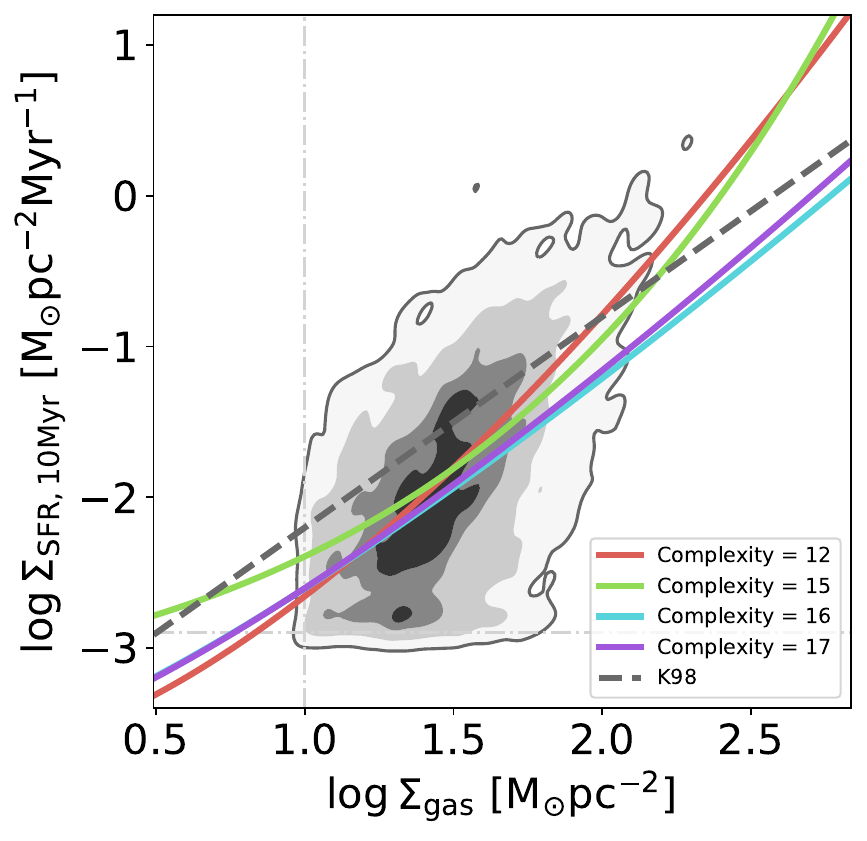}
    \hspace{0.5cm}
    \includegraphics[width=0.48\textwidth]{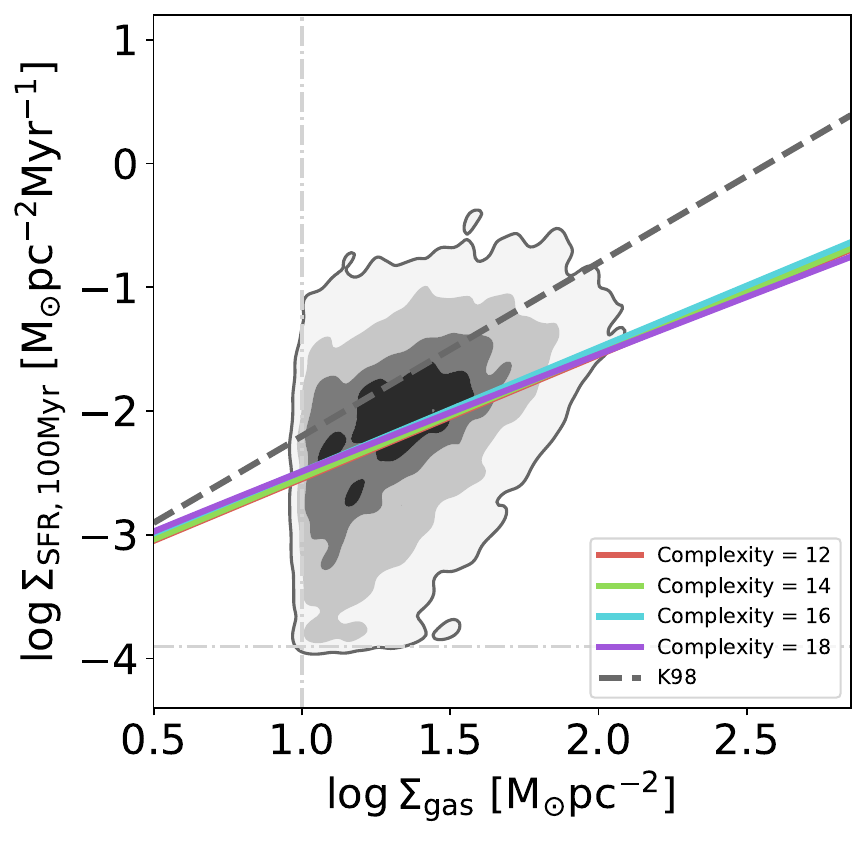}
    \caption{Distribution of points across the KS plane of the native FIRE data (grey contours), with the found models at fiducial values for features other than $\siggas$ overlaid, colored by complexity of the models. Style as Figure~\ref{fig:SFR_comparable_models}.}
    \label{fig:KSbehaviour_alleqns}
\end{figure*}

However, we can say that out of our two experiments, our \textsc{PySR}-based pipeline is better optimized to reach a stable solution for the $\tsfr$ = 100 Myr run, thus these results can be considered more reliable than those of its $\tsfr$ = 10 Myr counterpart. We can come to this conclusion because the models found for this experiment are all very similar, differing only in the values of the exponents and constants in the equations, and where there are indeed exponents they are very close to 1. As the models are all very similar, they exhibit very similar distributions within the feature space planes. Furthermore, represented by the pink line in the KS panels, we also show how the models would behave at fiducial values for the features besides $\siggas$ that come up in the models found, namely $\sigstars$, $\sigma_{\mathrm{gas,~z}}$ and $f_\mathrm{gas}$, for a local $L_*$ galaxy. All these models are summarized, color-coded by the complexity of each equation, in Figure~\ref{fig:KSbehaviour_alleqns}. Whilst there is some scatter in the equations found in the $\tsfr$=10 Myr experiment, it is clear that all the models found for the $\tsfr$= 100 Myr experiment converge to the same relation, once again reiterating that our pipeline has found the best possible model for this given dataset. At $\tsfr$=100 Myr timescales, the system is in more of a state of equilibrium because averaging over a longer time scale reduces stochasticity. Therefore, these results show that whilst our \textsc{PySR}-based pipeline perform well at producing predictive robust equations for similar near-equilibrium datasets, it struggles to do so at the same caliber for datasets encompassing lots of stochastic processes.

Finally, we consider the weighting of the gas velocity dispersion ($\gasvdisp$), on these top found equations, which through the unique characteristics on its projected distribution on the $\sigsfr$-$\gasvdisp$ feature plane we can infer to be an independent variable. The key characteristic of independent variables is that knowing the value of one does not affect the probability of the other, thus the distribution of points in the feature space plane of two independent variables would look like a 2D Gaussian distribution. We see from the middle columns of both Figures~\ref{fig:SFR10Myr_KSbehaviour} and ~\ref{fig:SFR100Myr_KSbehaviour}, that whilst there appears to be a loose correlation between $\gasvdisp$ and the the target variable, $\sigsfr$, the distribution in this feature space plane of both the native FIRE-2 data \textit{and} all the found models resembles a Gaussian distribution much more closely than that of $\siggas$ and $\sigstars$ feature space planes, especially when considering the contour within which 80$\%$ of the data lies. This indicates that $\gasvdisp$ is much more weakly coupled with $\siggas$ and $\sigstars$ thus is a more independent variable. This is supported by observations and simulations that also indicate that $\gasvdisp$ is weakly correlated with $\siggas$. Although $\siggas$ and $\sigstars$ have their highest densities toward the centers and are characterized by a factor of about 100 difference between the centers and outskirts \citep[e.g.,][]{AlataloHCG2015, Orr2018FIREKSlaw}, in comparison the radial profile of $\gasvdisp$ is relatively flat, being observed to have only about a factor of about 2 difference \citep[e.g.,][]{BattagliaEtAl2005, OrrEtAl2020SwirlsOfFIRE, KrumholzBurkhart2018GalacticDiscsModel}. This reiterates the findings shown by Figure~\ref{fig:train_metrics} in Subsection~\ref{subsec:eqn_performance} that $\siggas$ and $\sigstars$ are the predominant tracers of $\sigsfr$ in our dataset, with addition of $\gasvdisp$ to equations describing the target variable $\sigsfr$ as a secondarily influential variable added to fine-tune the fit of the model.

\section{Discussion} \label{sec:discussion}
In this study we have developed a methodology to find equations governing a given dataset in a data-driven manner. This procedure can be replicated and applied to any large dataset in order to give rise to symbolically-represented predictive models describing a target variable. The procedure does so by 1) establishing which of the many candidate input variables should qualify for training, and 2) applying the symbolic regression software \textsc{PySR} to attain equations that best describe the target variable by said selected input variables. In our star formation-specific problem, we are able to select for the the top four equations with an additional criteria that the equation generated by the algorithm must exhibit a final loss that is less than the loss of the best analytic equation tested in this study, as calculated by Equation~\ref{eq:LOSS_EQN}. However, in the general case this additional criteria is not strictly required for selecting an optimized equation; the default criteria based on the "score" of the equation could suffice. This study also presents a novel way to analyze the equations that are attained through the symbolic regression method, namely to not only look at the $R^2$ statistics but also explore the distribution of data across the feature space planes of the input variables that are prevalent in the final equations.

We find that our models are suggestive that the physics by which they were constructed are similar to that of many top-down approaches to parameterizing star formation. The \citet{FaucherGiguere13} and \citet{OstrikerShetty2011} models are equivalent when one considers a galactic disk in equilibrium such that the Toomre-Q parameter is constant and the gas fraction is 1, leading to a model where $\sigsfr \propto G\siggas^2$. This accounts for the weight of the ISM due to self-gravity. The \citet{OstrikerKim2022} model builds upon \citet{OstrikerShetty2011} to also account for the weight of the ISM due to stars, resulting in a model which is equivalent to a proportionality between $\sigsfr$ and the dynamical equilibrium pressure ($P_{\mathrm{DE}}$), defined as $P_{\mathrm{DE}} = \frac{\pi G}{2}\siggas^2 + \siggas\sqrt{2\rho_* \gasvdisp}$, where $\rho_*$ is the mid-plane stellar mass density expressed as $\rho_* = \sigstars/(4H_*)$, with $H_*$ being the disc scale height. The first term in the expression is that expressing the weight of the ISM due to self-gravity whilst the second is the weight of the ISM due to stellar gravity, for a distribution of gas and stars in a galaxy disk treated as isothermal fluids in a plane-parallel geometry. At the gas-rich limit where the weight of the stars are negligible compared to that of the gas, this model reduces to the underlying \citet{OstrikerShetty2011} model. It is of note that the variables $\siggas$ and $\sigstars$ appear in all of our found equations, and that at slightly higher complexities so does $\gasvdisp$. These variables also feature in the description of $P_{\mathrm{DE}}$, which has been observed to show strong correlation to the $\sigsfr$ in a dataset of 28 nearby galaxies in the the PHANGS-ALMA survey \citep{SunEtAl2020}, the TIGRESS simulations \citep{OstrikerKim2022}, and in the ALMaQUEST survey \citep{EllisonEtAl2024ALMaQUEST}, the latter of which found through a random forest analysis that $P_{\mathrm{DE}}$ was more influential in predicting target variable $\sigsfr$ than either the column density of molecular hydrogen, $\Sigma_{\mathrm{H_2}}$, a proxy for $\siggas$, or $\sigstars$, on their own. As the equations found in this study exhibit scaling relation-like relations indeed featuring all the variables featured in $P_{\mathrm{DE}}$, which feature prominently in top-down models, this is indicative that the models found via our procedure also resembles top-down equations.

Furthermore, as $\sigstars$ represents the weight of the bulk of the mass of the disk locally and thus can be considered a proxy for the potential of the disk, it is considered more of a "top-down" variable and is indeed does not play a factor in any "bottom-up" models. We can therefore once again infer that this study finds that the most apt description for star formation can be that of the "top-down" category. Even so, it is worth noting that the FIRE-2 pixel resolution of 750 parsecs that we use in our analysis does not resolve the giant molecular cloud (GMC) scale or the sub-parsec scales on which bottom-up models were constructed, thus it is unsurprising that top-down-like models dominate our \textsc{PySR}-found equations.

The pipeline and results presented in this study can be adapted for and applied to a plethora of theoretical galaxy evolution studies and methods. For example, while semi-analytic models (SAMs) for galaxy formation are powerful tools for understanding large-scale processes, they generally do not explicitly model the detailed vertical gas velocity distribution within galaxies. This distribution is crucial because it directly impacts the dynamics and evolution of the gas disk, which in turn influences star formation and feedback processes. Thus, its accurate representation allows models to better capture the effects of gas instabilities, regulate star formation rates, and predict the overall structure and evolution of galaxies. To fully capture this variety of processes, highly resolved simulations of the ISM that are more computationally expensive and only simulate a small region of a galaxy are required, so SAMs often estimate it instead by assuming axisymmetry and incorporating the radial distribution and velocity field of the gas, or by using empirical pressure-based recipes derived from observations to determine the distribution of gas in different phases \citep{CombesBecquaert1997, PoppingSomervilleTrager2014}. Substituting these approximated models in SAMs for a data-driven description of the vertical gas velocity distribution obtained through training SR on ISM simulations could therefore offer more physically nuanced constraints from which SAMs can predict galaxy formation. Likewise, the equations that were found by training on the FIRE-2 dataset in this study could directly replace the SF subgrid models of large-volume cosmological simulations to improve upon the empirically-based models that are instead traditionally called upon. 

\subsection{Caveats and Limitations}
We note the following caveats and limitations of our study:

\subsubsection{Capturing stochasticity}
 An observed limitation of our \textsc{PySR}-based pipeline is that the models it finds show difficulty capturing the same degree of stochasticity as the actual data at $\tsfr$=10 Myr timescales. The shorter time scale for star formation is inherently susceptible to a larger degree of stochasticity due its sensitivity to phenomena that occur on a shorter timescale, such as supernovae and other forms of stellar feedback. This is evident from the top equations yielded for this dataset's experiment, which show some degree of scatter within the KS plane when fiducial values of $\sigstars$ and $\gasvdisp$ are applied. The inconsistency amongst these equations found is shown in the left panel of Figure~\ref{fig:KSbehaviour_alleqns}. On the other hand, when applying this same metric to the models found $\tsfr$=100 Myr timescales, is it evident that all the relations converge to a line on the KS plane, evident in the right panel of Figure~\ref{fig:KSbehaviour_alleqns}. This indicates that this pipeline is already optimized to find the best possible predictive models at $\tsfr$=100 Myr timescales. This comparatively more robust model gives the appearance of a "top-down"-like scaling relation that related $\sigsfrHUNDREDMYR$ to $\siggas$, $\sigma_\mathrm{gas,~z}$ and $\sigstars$ for this dataset of FIRE-2 simulations. Yet because the models obtained when we train \textsc{XGBoost} - representative of the best possible model attainable with the variables provided - do not achieve very high $R^2$ values, the degree to which we can attain models with better statistical performance is constrained by inherent limitations in the dataset used in this study's experiments. As mentioned in Subsection~\ref{subsec:feature_space_planes}, our pixel-scale resolution of 750 pc does not resolve the GMC scale. Therefore, in order to develop a usable sub-grid model it would be best to replicate this study with a broader training dataset with finer resolution of at least 200pc, which is the upper limit of the GMC spatial scale \citep{Murray2011}. Follow-up studies could also apply the developed pipeline to data mapped onto a finer resolution Cartesian grid in order to investigate the resolution threshold, if any, at which smaller-scale physics might take over and resultant equations start to take more of a form reminiscent of "bottom-up" models.


\subsubsection{Radial profile of $\gasvdisp$}
We mention above that the radial profile of $\gasvdisp$ is generally found to be rather flat \citep[e.g.,][]{BattagliaEtAl2005, KrumholzBurkhart2018GalacticDiscsModel, OrrEtAl2020SwirlsOfFIRE}, thus $\gasvdisp$ and $\sigsfr$, which exhibits a much more dramatic change in density as one moves out from the center of the galaxy, would only harbor a weak correlation. This may be surprising because it is clear that SF drives turbulent motions of gas on small scales, as turbulence in the ISM is driven by an admixture of physical processes (e.g., supernova feedback, stellar winds, photoionization, gravitational accretions/radial flows). However, on scales larger than those of the individual supernova remnants, it is hard to infer a direct casual connection to the velocity dispersion seen. Adjacent areas within the galaxy thus interact with each other such that the gas velocity dispersion signals gets washed out, leading to a fairly constant signal across the galaxy. However, we also point out once again that since our pixel-scale resolution of 750 pc does not resolve the GMC scale, the correlation between $\gasvdisp$ and $\sigsfr$ may be even looser due to the coarse resolution of our dataset.

\subsubsection{Technical considerations}
Regarding the technical elements of our pipeline, a reconsideration of the training loss could be made. The addition of the quantile loss to the classical predictive loss did not result in a significant change in experimental results, as evident by the small contribution to the total loss of the quantile loss. This could be partly due to the choice of a suboptimal ratio of contribution to total loss of predictive and quantile loss, which would need to be determined by a grid search. However, this approach would be both computationally expensive and time-consuming. An alternative approach to optimize the loss function would be to change the choice of loss function altogether. For example, the Kullback-Leibler (KL) divergence measures the distance between the predicted probability distribution and that of the true probability distribution of the datasets being considered \citep{Csiszar1975, TervenEtAl2023, CiampiconiEtAl2023}. It is particularly relevant in this case of the variables explored in our study because, as mentioned in Subsection~\ref{subsec:feature_space_planes}, the resultant equations overestimate the strength of the correlation between the target variable and the most influential input variable, leading to a weak capture of the variance of the distribution of the original data in the feature space plane of said variable. The left-most column of Figures~\ref{fig:SFR10Myr_KSbehaviour} and right-most column of Figure~\ref{fig:SFR100Myr_KSbehaviour} most demonstrably captures this trend. The KL divergence loss could therefore be leveraged in subsequent studies to potentially lead to models that better capture the distribution of the true dataset.

\section{Conclusions} \label{sec:conclusions}
In this study, we applied a machine learning method called symbolic regression (SR) on the Latte suite of seven galaxies of the FIRE-2 galaxy simulations in order to produce an automated pipeline that yields a symbolic representation of star formation rate surface density ($\sigsfr$) from the data; an equation described by the most influential features out of a range of eight possible chosen input variables. As a point for comparison and benchmarking, we also calculate predictions for $\sigsfr$ as described by the ML decision-tree-based model XGBoost trained on the same input parameters, as well as a range of analytic SF models, including K98, three "bottom-up" models and a "top-down" model. Our method adopts a custom loss function based on the sum of the mean squared error loss and the quantile loss. The SR algorithm returns a collection of equations at different complexities, so to optimize for model selection, we require that top equations exhibit a final loss that is less than that of the best-performing analytic star formation model for that dataset. Of the equations that meet this criteria, four top equations are selected per experiment (see Table~\ref{tab:final_eqns} for found equations), ranked in order of highest achieving "score"; a metric that reflects the compromise between final loss and complexity of the equation.

We conducted two independent experiments, both trained to 4000 epochs; one with a target variable of the star formation rate column density averaged over 10 Myr, $\sigsfrTENMYR$, and that averaged over 100 Myr, $\sigsfrHUNDREDMYR$, trained in log space due to the large dynamic range of all the variables considered in this study. Through this exploration, we find that:
\begin{itemize}
    \item We have succeeded in developing a data-driven pipeline for the automated construction of equations that describe $\sigsfr$ that statistically outperforms the classic analytic star formation models assessed in this study. We analyze the behavior of these ML-found models through (1) taking note of the variables that feature most prominently in the algorithm-found equations and subsequently their construction within the equations, (2) calculating the goodness-of-fit $R^2$ of the found equations and studying the ensuing relation between $R^2$ and complexity of the equations, and (3) exploring the distribution of points as predicted by the algorithm-found equations compared to that of the native FIRE-2 simulations on the feature space planes of the variables that present most prominently in the equations.



    \item $\siggas$, $\sigstars$ and (although to a lesser extent,) $\sigma_{\mathrm{gas, z}}$, are required to construct a physically meaningful description for $\sigsfr$. This suggests that in addition to gas density, the potential of the ISM and turbulence-regulated motions within the ISM clouds are the most critical features to consider when constructing realistic descriptions of $\sigsfr$.

    \item Training on a target variable encompassing a longer timescale for SF, $\sigsfrHUNDREDMYR$, resulted in equations that all converge to a physically interpretable scaling relation akin to that attained by "top-down" approaches to parameterizing SF. This demonstrates that the pipeline proposed in this study successfully captures robust correlations between variables based on their physical interactions, as opposed to stochastic fluctuations, in systems closer to equilibrium.

    \item Our finding is consistent with the results from a SHAP analysis of an XGBoost model trained on the same parameters as that of the SR pipeline and is supportive of studies that find a tight correlation between $\sigsfr$ and the dynamical pressure equilibrium, $P_{\mathrm{DE}}$, which also features these variables. Simulations that do not resolve the vertical disk gas scale height may, therefore, be missing one of the important predictive variables for SF.

    \item The distribution of points across the feature space planes of the input variable that is most influential in describing the target variable appears to be very narrow when compared to the native FIRE-2 simulation distributions. This indicates that the models our pipeline finds slightly overestimate the strength of the correlation between the most influential input parameter and the target variable and, thus, do not fully capture the variance of the original data on which it was trained. Future work will aim to produce predictive models that more accurately capture the distribution of points across the feature space planes.
\end{itemize}

\begin{acknowledgments}
D.M.S. is supported by the 2022 Future Investigators in NASA Earth and Space Science and Technology (NASA FINESST) Fellowship (NASA grant 80NSSC22K1604). D.M.S. is also grateful for the generous support of the 2023 Quad Fellowship.
 B.B. acknowledges support from NSF grant AST-2009679 and NASA grant No. 80NSSC20K0500.
B.B. is grateful for the generous support of the David and Lucile Packard Foundation and the Alfred P. Sloan Foundation.
The Flatiron Institute is supported by
the Simons Foundation. All authors are grateful for the use of both GPU and CPU resources on the Rusty Supercomputer cluster at the Simons Foundation Flatiron Institute. All authors are deeply grateful for in-depth discussions with and suggestions from Viraj Pandya, Christian Kragh Jespersen, Chris McKee, Shyam Menon and Deanne Fisher.
\end{acknowledgments}

\appendix
\section{Derivation of bottom-up models} \label{appendix:bottom-up}


Expanding upon the K98 empirical scaling of $\sigsfr$ to $\siggas$, a yet tighter relation for $\sigsfr$ is achieved in \citet{KrumholzDekelMcKee2012} (hereafter KDM12) when the self-gravity of the immediate surroundings are also taken into account, described by a parametrization of $\Sigma_{\mathrm{SFR}}$ by the ratio between the mean column gas density and the mean free-fall time $\tff(\rho_0)$ of the cloud, which is the timescale required for a medium with negligible pressure support to collapse.
\begin{align}
\sigsfr =  \epsilon_{\mathrm{ff}}\frac{\siggas(\rho_0)}{\tff(\rho_0)} ~, \label{eq:KDM12}
\end{align}
where $\epsilon_{\mathrm{ff}}$ is the efficiency of the gas and the free-fall time of the gas, which is dependent on the gas density $\rho$ (with $\rho_0$ above being the mean gas density of the cloud), being defined by:
\begin{align}
    t_{\mathrm{ff}}(\rho) = \sqrt{\frac{3\pi}{32G\rho}}.
\end{align}\label{eq:tff}
Calibrating for Milky Way (MW) clouds and Local Group galaxies, this efficiency is $\approx 1\%$, and indeed achieves a stark improvement in star formation rate (SFR) correlation over the K98 relation \citep{Federrath2013sflaw, KrumholzReview2014}.

As touched on above, supersonic turbulent random motions in star-forming gas have also long been demonstrated both in observations and simulations to play a critical factor in regulating SF \citep{ZuckermanEvans1974, Larson1981, SolomonEtAl1987, FalgaronePugetPerault1992, HeyerBrunt2004, SchneiderEtAl2011}, with the force required to drive it locally thought to be originating simultaneously from within the cloud via stellar feedback as well as external to the cloud from shear or cloud-cloud interactions with its neighbors\citep{Federrath:2016, Federrath2017IAUsymposium}. The role of turbulence in regulating SF is dichotomous: on the one hand, the stabilization of molecular clouds on large scales by turbulent kinetic energy can prevent collapse and thus serve to suppress SF, yet, on the other hand, the local compressions in shocks induced by the turbulent fluid's supersonic nature in turn produce filaments, the intersections of which generate the required dense cores ripe for the initial conditions of SF \citep{MacLowKlessen2004, ElmegreenScalo2004, McKeeOstriker2007, SchneiderEtAl2012}. Such theories incorporating the turbulent supersonic motions are supported by consistency with observations of the initial mass function of stars \citep{PadoanNordlund2002, HennebelleChabrier2011, HennebelleChabrier2013} as well as that of star formation rates and efficiencies \citep{KrumholzMcKee2005, PadoanNordlund2011, HennebelleChabrier2011, FederrathKlessen2012, SalimEtAl2015, Salim2020HCG}. It is also well-established that the gas density closely follows a log-normal probability density function (PDF) confirmed by both observations and simulations \citep{KrumholzDekelMcKee2012, Federrath2013sflaw}:
\begin{align}
    p(s) = \frac{1}{\sqrt{2\pi\sigma_s^2}}\exp\Big(-\frac{(s-s_0)^2}{2\sigma_s^2}\Big)ds,
\end{align}
where $s$ is the natural logarithm of the mean-normalized density $\rho$:
\begin{align}
    s = \ln\Big(\frac{\rho}{\rho_0}\Big)
\end{align}
and $\sigma_s$ is the logarithmic density variance as derived by \citet{PadoanNordlund2011} and \citet{MolinaEtAl2012}:
\begin{align}
    \sigma_s^2 = \ln\Big(1+b^2\mathcal{M}^2\frac{\beta}{\beta + 1}\Big)^{3/8}.
\end{align}\label{eq:sigma_s_2}
$\sigma_s^2$ is characterized by the following turbulence-related properties: firstly, the turbulence driving parameter $b$, which indicates the degree of compressive versus solenoidal turbulence that exists within the system \citep{FederrathKlessenSchmidt2008, Federrath2010} with 1 representing completely compressive (curl-free) forcing, 1/3 being completely solenoidal (divergence-free) forcing and 0.4 taken to be the average mixed forcing. Next is the sonic Mach number $\mathcal{M}$ - the ratio between the absolute velocity of the gas, which in this study is estimated by the velocity dispersion of gas $\sigma$, and the local speed of sound $c_s$. Finally, we have the ratio of thermal to magnetic pressure, plasma $\beta$, a value of $\infty$ of which represents a lack of magnetic fields.

A \textit{multi-freefall} definition of SFR is attained when one considers that gas with high density $\rho$ exhibits shorter freefall times than diffuse gas, thus $\Sigma_{\mathrm{SFR}}$ can be parameterized by the integration of the entire gas density PDF, as in \citet{SalimEtAl2015} (hereafter SFK15):
\begin{align}
    \Sigma_{\mathrm{SFR}} &= \epsilon_{\mathrm{ff}} \cdot \int_{-\infty}^{\infty} \frac{\siggas}{\tff} \exp\Big(\frac{3}{2}s\Big)p(s)~ds \\
                          &= \epsilon_{\mathrm{ff}} \cdot \frac{\siggas}{\tff}\exp\Big(\frac{3}{8}\sigma_s^2\Big) \\
                          &= \epsilon_{\mathrm{ff}} \cdot \frac{\siggas}{\tff}\Big(1+b^2\mathcal{M}^2\frac{\beta}{\beta + 1}\Big). \label{eq:SFK15}
\end{align}
This relation has shown good direct correlation to $\Sigma_{\mathrm{SFR}}$ for the Small Magellanic Cloud (SMC) \citep{BolattoEtAl2011}, Central molecular Zone (CMZ) \citep{YusefZadeh2009}, numerous Galactic molecular clouds \citep{HeidermanEtAl2010, WuEtAl2010, GutermuthEtAl2011, LadaLombardiAlves2010} and a high-redshift object \citep{Sharda2018highzSFlaw}.

A further refinement of this method is attained when considering the gas density PDF integral from the critical density $s_{\mathrm{crit}}$, ensuring that only the gas that \textit{can} collapse to form stars, where $s_{\mathrm{crit}}$ is defined as that in \citet{KrumholzMcKee2005} and is based on comparing the Jeans length to the sonic length and extended by \citet{FederrathKlessen2012} to include magnetic fields by comparing $s_{\mathrm{crit}}$ to the magnetic Jeans length with the magneto-sonic scale:
\begin{align}
    s_{\mathrm{crit}} = \ln \Big[\Big(\frac{\pi^2}{5}\Big)\phi^2_x \alpha_{\mathrm{vir}}\mathcal{M}\Big].
\end{align}\label{s_crit}
Here, $\phi^2_x$ is a fixed "fudge factor" of order unity (introduced in \citet{KrumholzMcKee2005}; see \citet{FederrathKlessen2012} for details) and $\alpha_{\mathrm{vir}}$ is the virial parameter, which is the ratio of twice the kinetic energy to the gravitational energy, $\alpha_{\mathrm{vir}} = 2E_{\mathrm{kin}}/|E_{\mathrm{grav}}|$.

Integrating from this critical density, the extended multi-freefall model as derived in \citet{Salim2020HCG} is:
\begin{align}
        \Sigma_{\mathrm{SFR}} &= \epsilon_{\mathrm{ff}} \cdot \int_{s_{\mathrm{crit}}}^{\infty} \frac{\siggas}{\tff} \exp\Big(\frac{3}{2}s\Big)p(s)~ds \\
                              &= \epsilon_{\mathrm{ff}} \cdot \frac{\siggas}{\tff} \exp \Big(\frac{3}{8}~\sigma_s^2 \Big)\frac{1}{2} \Big[ 1+ \mathrm{erf}\Big(\frac{\sigma_s^2 - s_{\mathrm{crit}}}{\sqrt{2 \sigma_s^2}} \Big)\Big] \\
                              &=\epsilon_{\mathrm{ff}} \cdot \frac{1}{2}\frac{\siggas}{\tff}\Big(1+b^2\mathcal{M}^2\frac{\beta}{\beta + 1}\Big) \Big[ 1+ \mathrm{erf}\Big(\frac{\sigma_s^2 - s_{\mathrm{crit}}}{\sqrt{2 \sigma_s^2}} \Big)\Big] \label{eq:SAF20}
\end{align}
With an assumed fudge factor of $\phi_x = 0.17 \pm 0.02$ which was determined by fitting to a set of 34 numerical simulations of molecular clouds with resulting SF rates \citep{FederrathKlessen2012}, \citet{Salim2020HCG} found that this model aptly describes the SF across the Hickson compact group (HCG) galaxy NGC7674.

\section{Derivation of top-down models }
In this work we consider the \citet{FaucherGiguere13} (hereafter FG13) "feedback-regulated" model, which argues that the rate of feedback momentum injected by massive stars is balanced by the rate at which turbulence dissipates into the ISM. Momentum from turbulence, $P_\mathrm{turb} = \siggas\sigma_{\mathrm{gas}}$, decays across a dynamical time $\Omega_\mathrm{dyn}$ as:
\begin{align}
   \frac{dP_\mathrm{turb}}{dt} = -\frac{\Omega_\mathrm{dyn}}{2}P_\mathrm{turb} = -\frac{\siggas\sigma_{\mathrm{gas}}\Omega_\mathrm{dyn}}{2},
\end{align}\label{eq:P_turb_dot}
whereas feedback is injected at a rate of:
\begin{align}
    \frac{dP_{\mathrm{inj}}}{dt} = \Big(\frac{P_*}{m_*}\Big)\frac{d\Sigma_*}{dt}  = \Big(\frac{P_*}{m_*}\Big)\sigsfr,
\end{align}\label{eq:P_inj_dot}
where  $(P_*/m_*)$ is the momentum yield per stellar mass formed with a fiducial value of 3000 km s$^{-1}$. Assuming that the velocity dispersions in all cylindrical coordinates are roughly equal we can substitute $\sigma_{\mathrm{gas}}^2 = 3\sigma_{\mathrm{gas,z}}^2$, thus balancing the turbulence dissipation and feedback injection, we have:
\begin{align}
    \Sigma_{\mathrm{SFR}} = \frac{\sqrt{3}}{2} \frac{\Sigma_{\mathrm{gas}}\Omega_{\mathrm{dyn}}\sigma_{\mathrm{gas,z}}}{(P_*/m_*)} \; . \label{eq:FG13}
\end{align}

\section{SHAP values of main experiment} \label{appendix:shap}
\begin{figure*}
    \includegraphics[width=\linewidth]{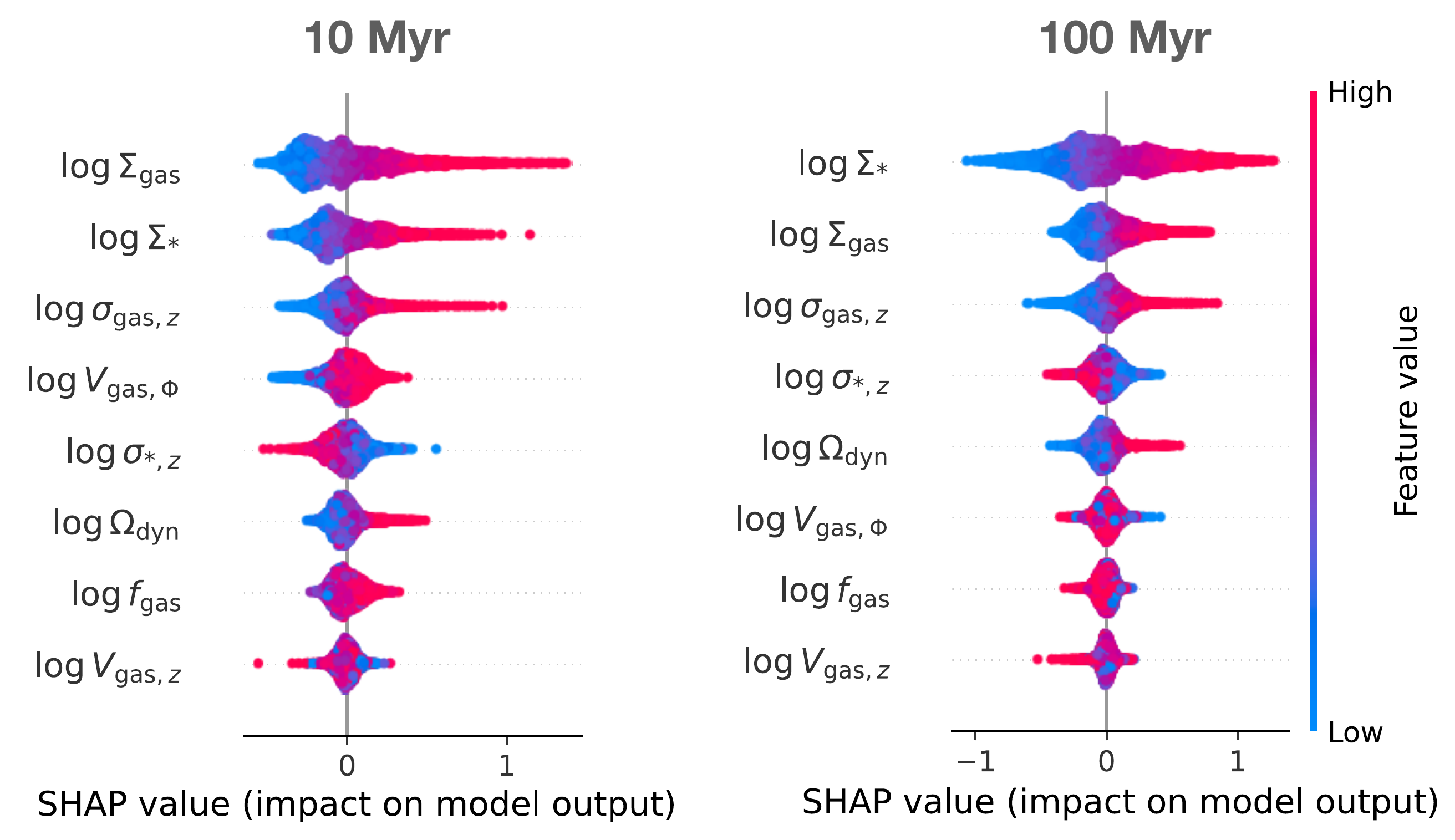}
    \caption{SHAP values for the input features used to train an XGBoost model for when SF is averaged over $\langle t_{\mathrm{SFR}} \rangle$ = 10 Myr (\textit{left}) and when averaged over $\langle t_{\mathrm{SFR}} \rangle$ = 100 Myr (\textit{right}). Each point in this figure indicates a the SHAP value of that corresponding data point, colored by the relative strength of that particular feature, with red indicating high values and blue indicating low values of that feature. A blue to red gradient across the beeswarm diagram for a particular feature indicates a correlation between that input variable and the target variable, whereas a red to blue gradient indicates an anti-correlation. Features are ranked from most predicted influence to least influence to the target variable from top to bottom respectively.}
\label{fig:shap_summary_plot}
\end{figure*}

 As an initial exploration into the degrees of influence the features in our data has on our target feature $\log\sigsfr$, we trained the scalable tree boosting regression system "XGBoost" \citep{Chen2016XGBoost} with a train to test split of $80\%$ to $20\%$ in order to calculate the SHAP (SHapley Additive exPlanations) values \citep{Lundberg2017SHAP} of each data point of each feature in this experiment, as shown in the "beeswarm diagram" shown in Figure~\ref{fig:shap_summary_plot}. (Note, we have trained everything in log scale due to the large dynamic range of all the features.) The SHAP value indicates the importance of a feature for a particular prediction, with values close to 0 indicating little to no influence and largely positive and negative SHAP values pointing to more influence in the high and low regimes of the target variable respectively. In this diagram features ranked from highest to lowest in predicted influence to the target variable are shown from top to bottom respectively. We observe that based on the calculated SHAP values, $\siggas$, $\sigstars$ and $\sigma_{\mathrm{gas,z}}$ are predicted to exert the most influence on $\sigsfr$ in both the $\tsfr$=10 Myr and $\tsfr$=100 Myr experiments.

\section{Generation of a synthetic FG13 model-based dataset}

\begin{figure*}
    \includegraphics[width=0.5\linewidth]{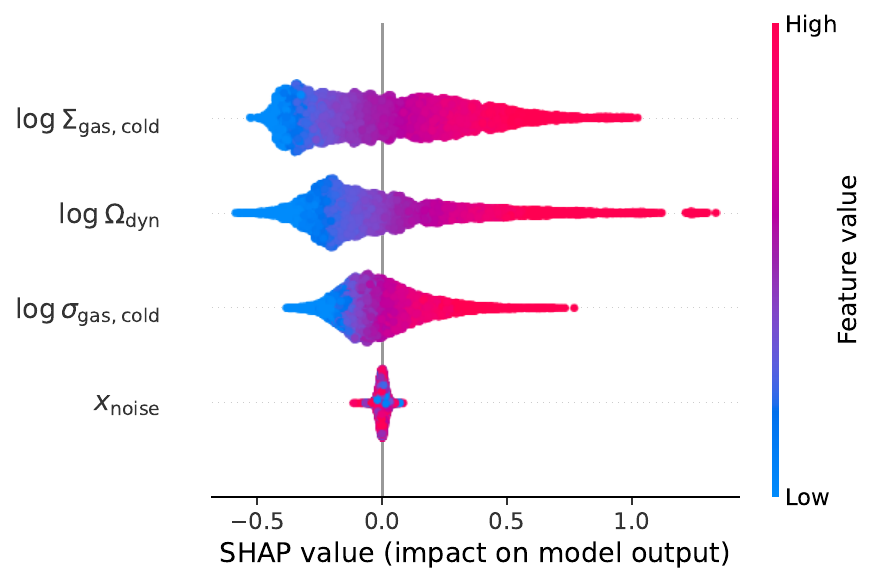}
    \caption{SHAP values for the four input features used to create the synthetic FG13 dataset and train an XGBoost model. Each point in this figure indicates a the SHAP value of that corresponding data point, colored by the relative strength of that particular feature, with red indicating high values and blue indicating low values of that feature. Features are ranked from most predicted influence to least influence to the target variable from top to bottom respectively.}
\label{fig:fake_fg13_shap}
\end{figure*}


To demonstrate the validity of our method of searching for equations we create a synthetic dataset based on the FG13 prescription of $\sigsfr$ averaged over 10 Myr as defined in Equation~\ref{eq:FG13}, where $\siggas$ and $\sigma_{\mathrm{gas}}$ are taken to be the cold dense gas and velocity dispersion of the cold dense gas in the FIRE-2 simulations respectively, thus denoted as $\Sigma_{\mathrm{gas,cold}}$ and $\sigma_{\mathrm{gas,cold}}$ for this experiment only. To simulate noise in the data, we add appropriate random noise to each variable. Random samples were drawn from a normal (Gaussian) distribution with means of 0 and standard deviations of 5$\mathrm{M_{\odot}pc^{-2}}$ for $\Sigma_{\mathrm{gas,cold}}$ which is about the transition density from HI to H$_2$, $10\%$ of the mean $\Omega_{\mathrm{dyn}}$ and 5~km s$^{-1}$ for $\sigma_{\mathrm{gas,cold}}$. We also created a pure noise variable $x_{\mathrm{noise}}$ by generating values from a random Gaussian distribution to test the resilience of our method to introduced features that are known to have no influence on the target feature.

We once again perform an initial probe into the degrees of influence the features in our data has on our target feature $\log\sigsfr$ by training the scalable tree boosting regression system "XGBoost" \citep{Chen2016XGBoost} with a train to test split of $80\%$ to $20\%$ in order to calculate the SHAP \ values \citep{Lundberg2017SHAP} of each data point of each feature in this experiment. The "beeswarm diagram" showing the prediction of the features exerting the most to least influence on the target variable is shown in Figure~\ref{fig:fake_fg13_shap}. It is encouraging that the pure noise variable $x_{\mathrm{noise}}$ exhibits SHAP values close to zero (i.e., no influence on $\log\sigsfr$) and is ranked the lowest in terms of its predicted capacity to influence our target variable $\log\sigsfr$. Furthermore, seeing the blue to red gradient on the beeswarm diagram indicates correlation between the feature and the predicted variable (with a red to blue gradient indicating anti-correlation). Here we can see that $\log\Sigma_{\mathrm{gas,~cold}}$ has an influential and consistent impact on the $\log\sigsfr$ prediction regardless of regime for the bulk of the range, but at the extreme high and low $\log\sigsfr$ regimes $\log\Omega_{\mathrm{dyn}}$ dominates impact, as exhibited by its very high and very low SHAP values at the highest and lowest feature values respectively.

We then applied the machine-learning (ML) based, symbolic regression (SR) equation search algorithm \textsc{PySR} \citep{Cranmer2023PySR} in order to explore the range of best equations that are representative of the dataset. As a genetic algorithm, equations are represented as a graph neural net, with the nodes in the graph exhibiting a term or operator in the equation. With each iteration, a phenomenon occurs to the graph akin to evolution after a generation, such as a random mutation whereby a node changes its operator, or a crossover, in which two separate branches within the graph can switch, thus coming up with new "generations" of equations. For more details regarding the algorithm and optimization methods of PySR please refer to \citet{Cranmer2023PySR}. In our experiments we constrain terms containing our input features to be linear combinations of each other (i.e., found equations can not contain direct products of input features) due to their logarithmic nature, and to improve interpretability in hopes of more easily identifying scaling relations. The final output of \textsc{PySR} are functional equations $\Psi$ predicting a relation between the target variable and the input variables. Therefore the final predictions of the target variable can be characterized as:
\begin{align}
   \widehat{\mathbf{y}} = \Psi(\mathbf{x}),
\end{align}
where for this experiment $\mathbf{y}=\sigsfr$ and
\begin{align}
    \mathbf{x}:=(\Sigma_{\mathrm{gas, cold}},~\Omega_{\mathrm{dyn}},~\sigma_{\mathrm{gas, cold}},~x_{\mathrm{noise}}).
\end{align}
For this initial experiment the prediction given by $\widehat{\mathbf{y}}$ of the found equation is compared to the values at the same location in the native FIRE simulation ${\mathbf{y}}$ via the standard mean squared error (MSE) loss. The MSE loss, also known as an $l_2$ regularized loss or prediction loss, is given by:
\begin{align}
  \mathcal{L}_{\mathcal{P}} = \dfrac{1}{N}\sum_{j=1}^{N}{\lr{\mathbf{y}_{j} - \widehat{\mathbf{y}}_{j}}\cdot\lr{\mathbf{y}_{j} - \widehat{\mathbf{y}}_{j}}} \label{eq:pred_loss_FG13}
\end{align}
where $N$ is the number of defined points in the simulations.

We define a "top equation" to be one that exhibits an MSE loss less than that of the best fit unity slope between $\Sigma_{\mathrm{SFR}}$ in the FIRE simulation and the $\Sigma_{\mathrm{SFR}}$ predicted from the FG13 model, ranked via the "score" goodness metric as detailed in \citet{Cranmer2023PySR}, which considers the loss of an equation against its complexity.

\begin{table}
\centering
\caption{PySR found equations for our generated data based on the FG13 prescription for $\sigsfr$.}
\begin{tabular}{llll}
    \hline
                     &       &       &         \\
    Model complexity & $R^2$ & Score & Equation\\
                     &       &       &         \\
    \hline
                     &       &       &         \\
    4                & 0.524 & 0.257 & $\log\Omega_{\mathrm{dyn}} - 3.40$ \\
    6                & 0.620 & 0.237 & $\log\Omega_{\mathrm{dyn}}^{1.38} - 3.62$ \\
    7                & 0.831 & 0.771 & $\log\Omega_{\mathrm{dyn}} + \log\Sigma_{\mathrm{gas, cold}} - 4.89$ \\
    \rowcolor{lightgray} 10               & 0.873 & 0.225 & $\log\sigma_{\mathrm{gas, cold}} + \log\Omega_{\mathrm{dyn}} + \log\Sigma_{\mathrm{gas, cold}} - 6.21$ \\
    \hline
\end{tabular}
\end{table}\label{tab:fake_fg13_eqns} 

The top four equations yielded from this method are tabulated in Table~\ref{tab:fake_fg13_eqns}. It is once again encouraging to observe that all top results for the parameterization of $\log\sigsfr$ are expressed in terms of the input features to the FG13 model, and that the pure noise variable was not included. $\log\Omega_{\mathrm{dyn}}$ unsurprisingly features very predominately in all the top equations found for this dataset, reflecting of the "top-down" nature of the FG13 model in describing SF in terms of the potential of the galaxy. Finally, it is very reassuring that the analytic form of the FG13 model (i.e., a linear combination of $\log\Sigma_{\mathrm{gas,~cold}}$, $\log\Omega_{\mathrm{dyn}}$ and $\log\sigma_{\mathrm{gas,~cold}}$) is achieved (shown in the shaded row in Table~\ref{tab:fake_fg13_eqns}), cementing the legitimacy of our equation search method for successfully recovering equations that accurately illustrate the inner workings of SF within galaxies.


\bibliography{SYMR_STARFORMATION}{}
\bibliographystyle{aasjournalv7}

\end{CJK*}
\end{document}